\documentclass[useAMS,usenatbib]{mn2e}
\usepackage{aas_macros}
\usepackage[dvips]{graphicx}
\usepackage{amssymb}
\title[AGN and Mass Distribution]{Mass Distribution in Galaxy Clusters: the Role of AGN Feedback}
\author[R. Teyssier et al.]{\parbox[t]{\textwidth}{Romain Teyssier$^{1,2}$\thanks{E-mail: romain.teyssier@cea.fr}, 
Ben Moore$^{1}$, Davide Martizzi$^{1}$, Yohan Dubois$^{3}$ and Lucio Mayer$^{1}$ }\vspace*{6pt}\\
$^{1}$Institute for Theoretical Physics, University of Zurich, CH-8057 Z\"urich, Switzerland\\
$^{2}$CEA Saclay, DSM/IRFU/SAP, B\^atiment 709, F-91191 Gif-sur-Yvette, Cedex, France\\
$^{3}$Astrophysics, University of Oxford, Keble Road, Oxford OX1 3RH, United Kingdom}

\begin{document}

\maketitle

\label{firstpage}

\begin{abstract}
We use $1\,{\rm kpc}$ resolution cosmological AMR simulations of a Virgo--like galaxy cluster to investigate
the effect of feedback from supermassive black holes (SMBH) on the mass distribution of dark matter, gas and stars. 
We compared three different models: (i) a standard galaxy formation model featuring gas cooling, star formation and supernovae feedback, 
(ii) a "quenching" model for which star formation is artificially suppressed in massive halos and finally (iii) the recently proposed AGN feedback model of \cite{Booth:2009p501}. Without AGN feedback (even in the quenching case), our simulated cluster
suffers from a strong overcooling problem, with a stellar mass fraction significantly above observed values in M87. The baryon distribution is highly concentrated, resulting in a strong adiabatic contraction (AC) of dark matter. 
With AGN feedback, on the contrary, the stellar mass in the bright central galaxy (BCG) lies below observational estimates and the overcooling problem disappears. The stellar mass of the BCG is seen to increase with increasing mass resolution, suggesting that our stellar masses converges to the correct value from below. The gas and total mass distributions are in better agreement with observations. We also find a slight deficit ($\sim$10\%) of baryons at the virial radius, due to the combined effect of AGN-driven convective motions in the inner parts and
shock waves in the outer regions, pushing gas to Mpc scales and beyond. This baryon deficit results in a slight {\it adiabatic expansion} of the dark matter distribution, that can be explained quantitatively by AC theory.   
\end{abstract}


\section{Introduction}

Galaxy clusters are ideal laboratories to study galaxy formation in a dense environment. Galaxies in clusters 
are observed in many morphological types, from blue extended spirals to red massive elliptical spheroids. The origin of
the morphological evolution of galaxies in clusters is still poorly understood. Tidal and ram pressure stripping
trigger fast evolution in the properties of accreted satellites, while complex gas cooling and heating processes
control the amount of stripped gas that can actually reach the central region of the cluster. In this context, the origin of 
bright cluster galaxies (BCG) still raises many questions. 
In the current cosmological framework, the formation of galaxies at the bright end of the
luminosity function is still affected by the so--called "overcooling problem": using both semi-anaytical models and computer
simulations, it has been shown that the massive galaxies are predicted to be too bright and too blue when compared to
massive galaxies in the nearby universe (see the recent review by \cite{Borgani:2009p728}). As a consequence, these models find a stellar content in massive clusters that
is significantly above the observed values \citep{Kravtsov:2005p702}, even including rather extreme supernovae feedback recipe \citep{Borgani:2004p1066}. 
 
In order to overcome this issue, feedback from supermassive black holes (SMBH) have been proposed as a mechanism to 
prevent gas from accumulating in the cluster core. The so--called AGN feedback scenario has received support from theoretical
considerations \citep{Tabor:1993p1080, Ciotti:1997p1087, Silk:1998p941}, but the strongest evidence comes from the correlated observations of X-ray cavities and radio blobs in massive 
clusters. These features are usually interpreted as buoyantly rising bubbles of high entropy material injected in the cluster core
by jets of relativistic particles. Detailed models of bubbles \citep{Churazov:2001p2079, Ruszkowski:2004p2156, Bruggen:2005p2162}  and jets  \citep{Reynolds:2001p2143, Omma:2004p2154, Cattaneo:2007p420} have been proposed in the context of cluster cores heating, usually based on spherically or planar symmetric,  idealized cluster configurations. 
Based on simple energetic arguments, it is possible
to relate the growth of SMBHs at the centre of massive galaxy spheroids to the energy required to unbind the overcooling gas 
\citep{Silk:1998p941}. The idea of star formation being regulated by AGN feedback at the high mass end of the galaxy mass function
has been applied first quite successfully to hydrodynamical simulations of galaxy merger \citep{DiMatteo:2005p1020}  and then to semi-analytical models of galaxy formation \citep{Croton:2006p5348, Bower:2006p1007, Cattaneo:2006p990, DeLucia:2007p1656, Cattaneo:2009p946}.

AGN feedback models have been included only recently in cosmological simulations of galaxy groups and clusters \citep{Sijacki:2007p1032, Puchwein:2008p767,McCarthy:2009p747}. Although the detailed physical modeling of SMBHs growth usually differ \citep{Sijacki:2007p1032, Booth:2009p501}, these simulations, exclusively based on the GADGET code \citep{Springel:2005p1064}, have obtained 
very encouraging results regarding the global properties of the simulated clusters \citep{Puchwein:2008p767, McCarthy:2009p747, Puchwein:2010p763, Fabjan:2010p787}.  In this paper, we report on the first AMR high-resolution simulation of a Virgo-like cluster, following both SMBH and star formation, 
with the associated feedback. A companion paper is exploring the properties of a jet--based, AGN feedback model \citep{Dubois:2010p5834}. We use the AMR code RAMSES, which differs significantly from the other code used so far to 
model AGN feedback in a fully cosmological simulation, namely the GADGET code. Although AMR schemes suffer from larger advection errors than SPH, which is a strictly Galilean invariant method,  there are some definitive advantages of using AMR in this context: although the total energy (kinetic + thermal + potential) is conserved only at the percent level, it is strictly conserving for the fluid energy (kinetic + thermal), which is very important in presence of strong shocks, and it captures hydrodynamical instabilities more realistically \citep{Agertz:2007p415, Wadsley:2008p1093, Mitchell:2009p4163}, which is very important in presence of convective motions of buoyant gas. It also captures gas stripping of infalling satellites due to hydrodynamical instabilities more realistically than standard implementations of SPH \citep{Agertz:2007p415}.

We have adapted the SMBH growth model of \cite{Booth:2009p501} to the sink particle method for AMR presented by \cite{Krumholz:2004p1079}. With respect to the previous work of \cite{Booth:2009p501} and follow-up papers, we have made significant improvements over the OWL simulations suites in term of mass and spatial resolution, but only for one single zoom-in simulation of a Virgo--like cluster. Recently, \cite{Puchwein:2008p767} and  \cite{Puchwein:2010p763} have also used the GADGET code, but a different AGN feedback model, to perform zoom-in simulations of groups and clusters of galaxies, with however a mass resolution and a gravitational softening length not as good as the one we used in our high resolution case. Note also that the minimum smoothing length in these SPH simulations is usually much smaller than the gravitational softening length. In this paper, we have specifically chosen a Virgo-like cluster, in order to compare our results with the very detailed observations that are available for this well-known astronomical object. We will focus here on the mass distribution of the three main components, namely dark matter, gas and stars.

The paper is organized as follows: the first section is dedicated to numerical methods and physical ingredient (cooling, star formation and AGN feedback), while our second section presents our results, comparing our three models. The final section is left for discussion.

\section{Numerical Methods}
\label{sec:num_methods}

In this section, we describe the numerical techniques and the initial conditions we used to model our Virgo--like cluster. 
As it is now customary for cosmological simulations, we used a zoom-in (or volume renormalisation) technique to focus
our computational resources on a specific region of a 100 $Mpc/h$ periodic box. We adopt a standard $\Lambda$CDM
cosmology with $\Omega_m=0.3$, $\Omega_\Lambda=0.7$, $\Omega_b=0.045$ and $H_0=70$ km/s/Mpc. We have 
adopted the \cite{Eisenstein:1998p1104} transfer function and the {\ttfamily grafic} 
package \citep{Bertschinger:2001p1123} in its parallel implementation {\ttfamily mpgrafic} \citep{Prunet:2008p388} to generate our initial conditions. From a first low resolution dark matter only run, we built a catalog of candidate halos whose virial masses lie in the range $10^{14}$ to $2 \times 10^{14}$ M$_\odot$/h. From this catalog, we have extracted our final halo based on its mass assembly history: its final mass ($M_{vir} \simeq 10^{14}$ M$_\odot$) is already in place around $z=1$, so it can be considered as a well relaxed cluster by $z=0$.  The final halo mass has been measured to be $M_{200c}=1.04 \times 10^{14}$~M$_\odot$ or $M_{500c}=7.80 \times 10^{13}$~M$_\odot$, where indice $c$ refers to the critical density. 

\subsection{Simulation parameters}

We have performed two simulations, one at low resolution, for which the initial grid effective size was $1024^3$, and
one at high resolution, with effective grid size of  $2048^3$. From this initial grid, we have extracted high resolution particles
only in the Lagrangian volume of the halo, and we have used larger and larger mass particles to sample the cosmological volume, 
so that the total number of particles in the box was $5.2 \times 10^6$ (resp. $22\times 10^6$) for only $2.6\times 10^6$ (resp. $19\times 10^6$) in the central region, and $10^6$ (resp. $8\times 10^6$) inside the final virial radius for the low resolution (resp. high resolution) simulation. The dark matter particle mass is therefore $6.5\times 10^7$ (resp. $8.2 \times 10^6$) M$_\odot$/h and the baryons resolution element mass is $1.2\times 10^7$ (resp. $1.4 \times 10^6$) M$_\odot$/h. 

The AMR grid was initially refined to the same level of refinement than the particle grid ($1024^3$ and $2048^3$), and
7 more levels of refinements were considered. We imposed the spatial resolution to remain roughly constant in physical units,
so that the minimum grid cell stayed close to $\Delta x_{\rm min} = L/2^{\ell_{\rm max}}$ with $\ell_{\rm max}=$17 (resp. 18) at $z=0$. We therefore
reached a spatial resolution of $\Delta x_{\rm min} \simeq 1$ kpc for the low resolution simulation and $\Delta x_{\rm min} \simeq 500$ pc
for the high resolution one. The grid was dynamically refined up to this resolution using a quasi-Lagrangian strategy: when the dark matter or baryons mass in a cell reaches 8 times the initial mass resolution, it is split into 8 children cells. We reached $z=0$ with $14 \times 10^6$ (resp. $68 \times 10^6$) cells for the low (resp. high) resolution run, including split cells.

Gas dynamics is modeled using a second-order unsplit Godunov scheme \citep{Teyssier:2002p451, Teyssier:2006p413, Fromang:2006p400} based on the HLLC Riemann solver \citep{Toro:1994p1151}. We assume a perfect gas equation of state (EOS) with $\gamma=5/3$. Gas metallicity is advected as a passive scalar, and is self-consistently accounted for in the cooling function. We also considered the standard homogeneous UV background of \cite{Haardt:1996p1167}, but we modified the starting redshift, extrapolating the average intensity from $z_{\rm reion}=6$ to $z_{\rm reion}=12$, in order to account for the early reionization expected in such a proto-cluster regions \citep{Iliev:2008p1200}.

\subsection{Galaxy formation physics}

\begin{table}
\begin{center}
\begin{tabular}{|l|c|c|c|c|}
\hline
Run & $m_{\rm cdm}$&  $m_{\rm gas}$ & $m_*$ & $\Delta x_{\rm min}$ \\
\hline
& & in $10^{6}$ M$_\odot$ & & in kpc/h \\
\hline
Low res & $65$ & $12$ & $2.4 $ & 0.76 \\
High res & $8.2$ & $1.4$ & $0.3$ & 0.38 \\
\hline
\end{tabular}
\end{center}
\caption{Mass resolution for dark matter particles, gas cells and star particles, and spatial resolution (in physical units) for our 2 sets of simulations. }
\end{table}

As gas cools down and settles into centrifugally supported discs, we need to provide a realistic model for the interstellar medium
(ISM). Since the ISM is inherently multiphase and highly turbulent, it is beyond the scope of present-day cosmological simulations to try to simulate it self-consistently. It is customary to rely on subgrid models, providing an effective EOS that capture the basic turbulent
and thermal properties of this gas. Models with various degrees of complexity have been proposed in the literature \citep{Yepes:1997p1245, Springel:2003p1288, Schaye:2008p1393}. We follow the simple approach based on a temperature floor given by a polytropic EOS for gas 
\begin{equation}
T_{floor} = T_* \left( \frac{n_{\rm H}}{n_*} \right) ^{\Gamma -1}
\label{eq:eos}
\end{equation}
where $n_*=0.1$ H/cc is the density threshold that defines the star forming gas, $T_*=10^4$ K is a typical temperature mimicking
both thermal and turbulent motions in the ISM and $\Gamma=5/3$ is the polytropic index controlling the stiffness of the EOS. Gas is able to heat above this floor, but cannot cool down below it. Note that because of this temperature floor, the Jeans length in our galactic disc is always resolved. We also consider star formation using a similar phenomenological approach. In each cell with gas density larger than $n_*$, we spawn new star particles at a rate given by
\begin{equation}
\dot \rho_{*} = \epsilon_* \frac{\rho_{gas}}{t_{\rm ff}}~~~{\rm with}~~~t_{\rm ff}=\sqrt{\frac{3\pi}{32G\rho}}
\end{equation}
where $t_{\rm ff}$ is the free-fall time of the gaseous component and $\epsilon_*=0.01$ is the star formation efficiency. The star particle
mass depends on the resolution and was chosen to be $2.4\times 10^6$ (resp. $3\times10^5$) M$_\odot$ for the low (resp. high) resolution run. For each star particle, we assume that 10\% of its mass will go supernova after 10 Myr. We consider a supernova energy of $10^{51}$ erg and one M$_\odot$ of ejected metals per 10 M$_\odot$ average progenitor mass. This supernovae feedback was implemented in the RAMSES code using the "delayed cooling" scheme \citep{Stinson:2006p1402}. 

Up to this point, we use rather standard galaxy formation recipe, which have proven only recently to be quite successful in reproducing the properties of field spirals \citep{Mayer:2008p1478, Governato:2009p1455, Governato:2010p1442, Agertz:2010p5352}.  These recipe have been shown to reproduce basic galaxy properties like Kennicutt-Schmidt law, star formation rates, galactic winds \citep{Dubois:2008p393, Devriendt:2010p5266, Agertz:2010p5352}. The same recipe are only marginally successful when one considers small groups \citep{Feldmann:2010p1516}, but they fail on cluster scales \citep{Borgani:2004p1066, Kravtsov:2005p702, Borgani:2009p728}. In the present paper, in order to check that our results are compatible with previous work, and to set a reference point, we have performed simulations with only standard galaxy formation physics (labelled SF runs in the followings). 

\subsection{Star formation quenching}

The main problem we have to face in standard cosmological simulations on cluster scales is the striking excess of mass locked into stars. Using rather extreme stellar feedback models, previous authors report that the simulated stellar mass fraction lies in the range 35 -- 60\%, depending on the cluster mass \citep{Borgani:2009p728}. Since only 10\% of the baryonic mass is observed in the galaxies, this would require a large amount of stars hidden in a diffuse component such as the intracluster light (ICL). This last scenario is however severely constrained by observations of the ICL \citep{Gonzalez:2007p916} and does not appear to be plausible.

There is growing theoretical and observational evidence that star formation is shutdown above a critical halo mass $M_c \simeq 6 \times 10^{11}$ M$_\odot $\citep{Cattaneo:2006p990}. \cite{Birnboim:2003p1535} have proposed that this critical mass is related to the stability of accretion shocks, and to a transition from cold to hot mode of gas accretion. Although this transition in the nature of the accretion flows has been confirmed by numerical simulations \citep{Keres:2005p1546, Ocvirk:2008p387, Dekel:2009p100}, the simulated star formation was not observed to decrease significantly above the critical mass \citep{Ocvirk:2008p387}. This might be due to insufficient mass and spatial resolution, so that heating processes, not properly captured, cannot balance radiative cooling \citep{Naab:2007p3844}. On a different side, \cite{Cattaneo:2006p990} argued that this modification of the halo properties may create favorable conditions for AGN feedback to be effective, but AGN feedback is still needed to prevent the overcooling problem above the critical mass \citep{Dekel:2006p3551}. Without specifying the underlying heating process, the critical mass argument boils down to stopping (or quenching) gas cooling and star formation in the central galaxy for halo masses larger than $\sim 10^{12}$ M$_\odot$.

In this paper, in order to test this hypothesis, we have used a simple phenomenological model to quench star formation in massive galaxies. Since our standard model (SF runs) obviously suffers from a strong overcooling problem, we need to actively suppress gas cooling and star formation above the critical mass. If the stellar mass density is greater than 0.1 H/cc, and if the stellar 3D velocity dispersion is greater than 100 km/s, we switch off star formation and gas cooling. In this way, star formation in discs is unaffected, since the velocity dispersion is smaller than the chosen threshold. On the other hand, star formation is suppressed in massive spheroids. The main advantage of this quenching model is its simplicity: although it captures the scenario proposed by \cite{Cattaneo:2006p990}, it does not require any complex AGN feedback model, nor the mass and spatial resolution reached by \cite{Naab:2007p3844} in their early--type galaxy simulation. However, as we demonstrate in this paper, this simple approach doesn't solve the overcooling problem in our Virgo cluster simulation.

\subsection{SMBH growth and associated feedback}

Beside our standard galaxy formation and our quenching scenario simulations, we explore a model for which SMBH growth and feedback is considered. In a nut shell, our recipe is based on the sink particle method for grid-based codes designed by \cite{Krumholz:2004p1079}, with the AGN feedback model proposed by \cite{Booth:2009p501}. We shall now briefly summarize these techniques.

\subsubsection{Seed particles}

In the two main competing SMBH formation models: slow growth from Pop III stars \citep{Madau:2001p3493} or direct collapse of a low angular momentum halo \citep{Bromm:2003p3495, Begelman:2006p3499}, the seed SMBH is believed to grow quickly to $10^5$  M$_\odot$, before starting to affect its environment and enter the self-regulated regime. Although the question of intermediate mass black holes is still vigorously debated, this characteristic mass is often considered as the initial seed SMBH mass \citep{Li:2006p3423, Pelupessy:2007p3492}, since it is one order of magnitude smaller than the smallest SMBH observed on the $M_{\rm BH}-\sigma$ relation \citep{Gebhardt:2000p1011, Gultekin:2009p475}. At each main time step during the course of the simulation, we identify potential candidate regions for seed black hole formation using the following criteria:
\begin{itemize}
\item the stellar density has to be greater than 0.1 H/cc
\item the stellar 3D velocity dispersion has to be greater than 100 km/s
\item the gas density has to be greater than 1 H/cc
\item no other sink particle is present within 10 kpc
\end{itemize}   
If these four conditions are fulfilled, we create a sink particle of fixed mass $M_{\rm BH}=10^5$ M$_\odot$. These particles represent our seed black holes.
Note that our third condition ensures that seed black holes form in the nuclear region of star forming disks, where the gas density is probably much larger than 1~H/cc. Because our limited resolution, we cannot choose an arbitrary high density threshold, otherwise SMBH will never form. On the other hand, because star formation is a very inefficient process, large enough galaxies devoid of SMBH will always reach this gas density threshold and trigger SMBH seeding.  Our second condition requires a minimum line--of--sight velocity dispersion $\sigma_{1D} \simeq 60$~km/s, in agreement with the observed $M_{BH}-\sigma$ relation \citep{Gebhardt:2000p1011, Gultekin:2009p475}. Our last condition ensures that no new seed SMBH will be created in a galaxy that is already hosting an old SMBH (or at least within 10 kpc from its center).

\subsubsection{Sink particle evolution}

Each black hole is considered as a sink particle, as defined in \cite{Krumholz:2004p1079}. We recall briefly the method here.
We consider around each black hole a sphere of fixed physical radius $r_{\rm sink}=4 \Delta x$, where $\Delta x$ is our spatial resolution in physical units, so that $r_{\rm sink} \simeq 4$ kpc (resp. $2$ kpc) for the low (resp. high) resolution run. We assume that the SMBH mass distribution inside the sink is homogeneous, and this homogeneous sphere is added to the total mass density when solving for the Poisson equation. The sink particle is then advanced in time by interpolating the gravitational force back to the sink position using the inverse CIC scheme. For each sink, we compute the Bondi-Hoyle accretion rate
\begin{equation}
\label{bondi_formula}
\dot{M}_{\rm BH}=\alpha_{\rm boost} \frac{4\pi{\rm G}^2M_{\rm BH}^2 \rho}{ (c_{\rm s}^2+u^2)^{3/2}}
\end{equation}
where $\rho$, $c_s$ and $u$ are the average gas density, sound speed and relative velocity within the sink radius. These average quantities are computed following the scheme proposed by \cite{Krumholz:2004p1079}. The parameter $\alpha_{\rm boost}$ was introduced by \cite{Springel:2005p63} to account for unresolved multiphase turbulence in the SMBH environment. Although this parameter was first chosen constant at $\alpha_{\rm boost} \simeq 100$  \citep{Springel:2005p63},  \cite{Booth:2009p501} argued that this parameter should be close to one in low density regions, while its value should increase in high density regions, in order to match the subgrid model used for the unresolved turbulence in the disks. Based on extensive numerical experiments, they proposed the following phenomenological model to boost Bondi-Hoyle accretion.
\begin{eqnarray}
\nonumber
\alpha_{\rm boost}&=&\left( \frac{n_{\rm H}}{n_*} \right)^2~~~{\rm if }~~Ên_{\rm H} > n_* = 0.1~{\rm H/cc,}\\
\alpha_{\rm boost}&=&1~~~{\rm otherwise.}
\end{eqnarray}
Note that this model has no real physical justification, and that it depends crucially on the underlying EOS model. We have been very careful in using this boost factor in conjunction with the same EOS model (see Equ.~\ref{eq:eos}) as in \cite{Booth:2009p501}. 

As advocated by \cite{Springel:2005p63}, the accretion rate onto the SMBH cannot exceed its Eddington limit given by
\begin{equation}
\label{eddington_formula}
\dot{M}_{\rm ED}=\frac{4\pi{\rm G}M_{\rm BH} m_{\rm p}}{\epsilon_r \sigma_{\rm T}c}~~~Ê{\rm with}~~\epsilon_r \simeq 0.1{\rm ,}
\end{equation}
so that the final accretion rate is computed using $\dot{M}_{\rm acc}=\min (\dot{M}_{\rm BH},  \dot{M}_{\rm ED})$. At each time step, a total gas mass of $\dot{M}_{\rm acc} \Delta t$ is removed from all cells within the sink radius, with the same weighting scheme as the one used to define average quantities \citep{Krumholz:2004p1079}. In order to prevent the gas density to vanish or become negative, we allow a maximum of 50\% gas removal at each time step.

\subsubsection{AGN feedback}
 
\begin{figure}
  \begin{center}
    \includegraphics[width=0.45\textwidth]{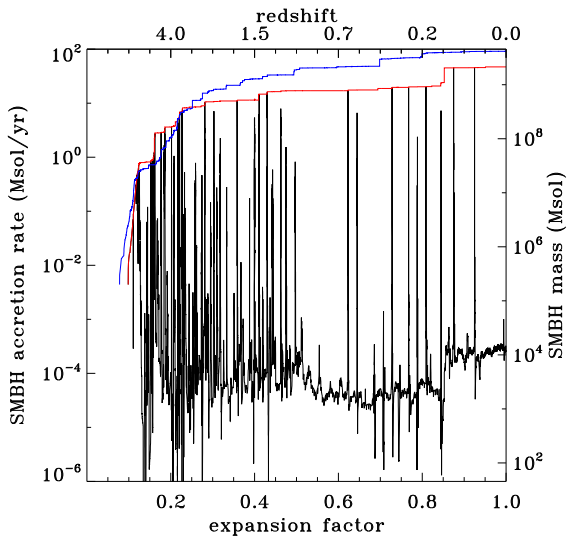}\\
  \end{center}
  \caption{ \label{fig:SmbhAcc} Time evolution of the accretion rate of the largest SMBH in the cluster. The Eddington limit is shown as the red line, the corresponding SMBH mass indicated on the right axis. The high resolution simulation finds a similar (within a factor of 2) SMBH mass (shown in blue). We clearly see two modes of accretion, with episodic bursts for which the SMBH accretes at or close to the Eddington limit, and long periods in between for which the SMBH accretion rate is at or close to the Bondi rate.}
  \label{fig:AgnGrowth}
\end{figure}

In the proposed model for SMBH growth, a key ingredient is the associated feedback model. As demonstrated by many authors \citep{Sijacki:2007p1032, Cattaneo:2007p420, Booth:2009p501}, the Bondi-Hoyle accretion model allows the SMBH growth history to be self-regulated by injecting thermal energy in the surrounding gas: if the SMBH mass is too small, the associated feedback will be inefficient and the surrounding gas will remain cold and dense, boosting the accretion rate up to the Eddington limit. The SMBH mass will grow exponentially fast, with e-folding time scale equal to the Salpeter time $t_{\rm S} \simeq 45$ Myr. As soon as the black hole mass is large enough, feedback processes heat and eventually unbind the surrounding gas, so that the accretion rate drops and becomes Bondi-Hoyle limited. This bimodal behavior is illustrated in Figure~\ref{fig:SmbhAcc}, where one can see the time evolution of the accretion rate of the most massive SMBH in the simulated region. Short bursts of Eddington limited accretion are followed by long quiescent epochs of Bondi-Hoyle limited accretion, self-regulated by SMBH feedback. 

In order to allow for this self-regulated SMBH growth, efficient feedback schemes are mandatory. Although we do see clear signatures of AGN feedback in clusters \citep{Arnaud:1984p2173, Fabian:2000p2204, McNamara:2001p2223}, the physical processes at the origin of this energy injection are still unclear: radiative feedback \citep{Ciotti:2001p2231}, cosmic rays \citep{Bruggen:2002p2160, Chandran:2007p2323} or strong shocks (see the review of \cite{Begelman:2004p2233}). A common property of these various models is that they require a very good spatial resolution to be captured realistically in hydrodynamical simulations. The most advanced modeling so far has been using AGN-driven bubbles \citep{Churazov:2001p2079, Ruszkowski:2004p2156, Bruggen:2005p2162} or jets \citep{Reynolds:2001p2143, Omma:2004p2154, Cattaneo:2007p420}, leading to turbulent convective motions and "shocklets" escaping the cluster core \citep{Chandran:2007p2323, Rasera:2008p2324, Sharma:2009p2272}. In some case, depending on the injected energy, these AGN-driven flows can drive strong shock waves that can travel to very large distances \citep{Baldi:2009p2344}. In current cosmological simulations,  numerical resolution does not allow these effects to be self-consistently modeled. As it is customary under these circumstances, we rely on a more phenomenological model.

In the cosmological context, the model proposed by \cite{Booth:2009p501} appears to be easier to implement than the one proposed by \cite{Sijacki:2007p1032}. Moreover, it relies on only one main free parameter, the coupling  efficiency $\epsilon_{\rm c}$, that can be calibrated to the observed $M_{{\rm BH}}-\sigma$ relation to the fiducial value $\epsilon_c \simeq 0.15$ \citep{Booth:2009p501}.
We have adapted their model to the RAMSES code, using the following approach: at each time step, we compute the SMBH feedback energy as a fixed fraction of the rest mass energy of the accreted gas, multiplied by the "coupling efficiency" $\epsilon_{\rm c}$,
\begin{equation}
\Delta E = \epsilon_{\rm c} \epsilon_{\rm r}\dot{M}_{\rm acc}c^2 \Delta t {\rm .}
\label{enerdump}
\end{equation}
This energy is not released immediately in the surrounding gas. It is instead accumulated over many time steps and stored into a new SMBH related variable $E_{\rm AGN}$, so that we can avoid atomic line cooling to radiate this energy instantaneously. Following the trick proposed by \cite{Booth:2009p501}, we release this accumulated energy inside the sink radius when 
\begin{equation}
E_{\rm AGN} > \frac{3}{2} m_{\rm gas} k_B T_{\rm min}
\end{equation}
where $m_{\rm gas}$ is the gas mass within the sink radius and $T_{\rm min}$ is the minimum feedback temperature. As soon as $T_{\rm min}$ is chosen above $10^7$ K, the critical temperature below which metal line cooling becomes very efficient, the resulting feedback scheme does not depend on the chosen value for $T_{\rm min}$. We adopt here $T_{\rm min}=10^7$ K. As can be seen on the previous equation, this threshold energy depends directly on the gas density in the environment of the black hole. When dense and cold gas is present, more energy is required to reach the threshold. After enough mass has been accreted, a strong burst of energy is released, that will unbind the surrounding dense gas. On the other hand, when only diffuse, hot gas is present, the threshold energy is much easier to reach, and feedback proceeds in a quasi-continuous fashion. In some sense, based on this rather simple recipe, we can account for both the "quasar mode" and the "radio mode" of the AGN feedback model of \cite{Sijacki:2007p1032}. 

\section{Results}
\label{sec:results}

In this section, we describe the properties of our simulated cluster for the 3 different models, labelled "SF", "quenching" and "AGN" in most of the Figures. We will focus our analysis in the final mass distribution, and compare, whenever it is possible, to actual Virgo cluster data. We present mostly low resolution data, although we also compare low and high resolution results to discuss convergence properties.

\subsection{SMBH Growth and Feedback}

\begin{figure*}
    \includegraphics[width=0.45\textwidth]{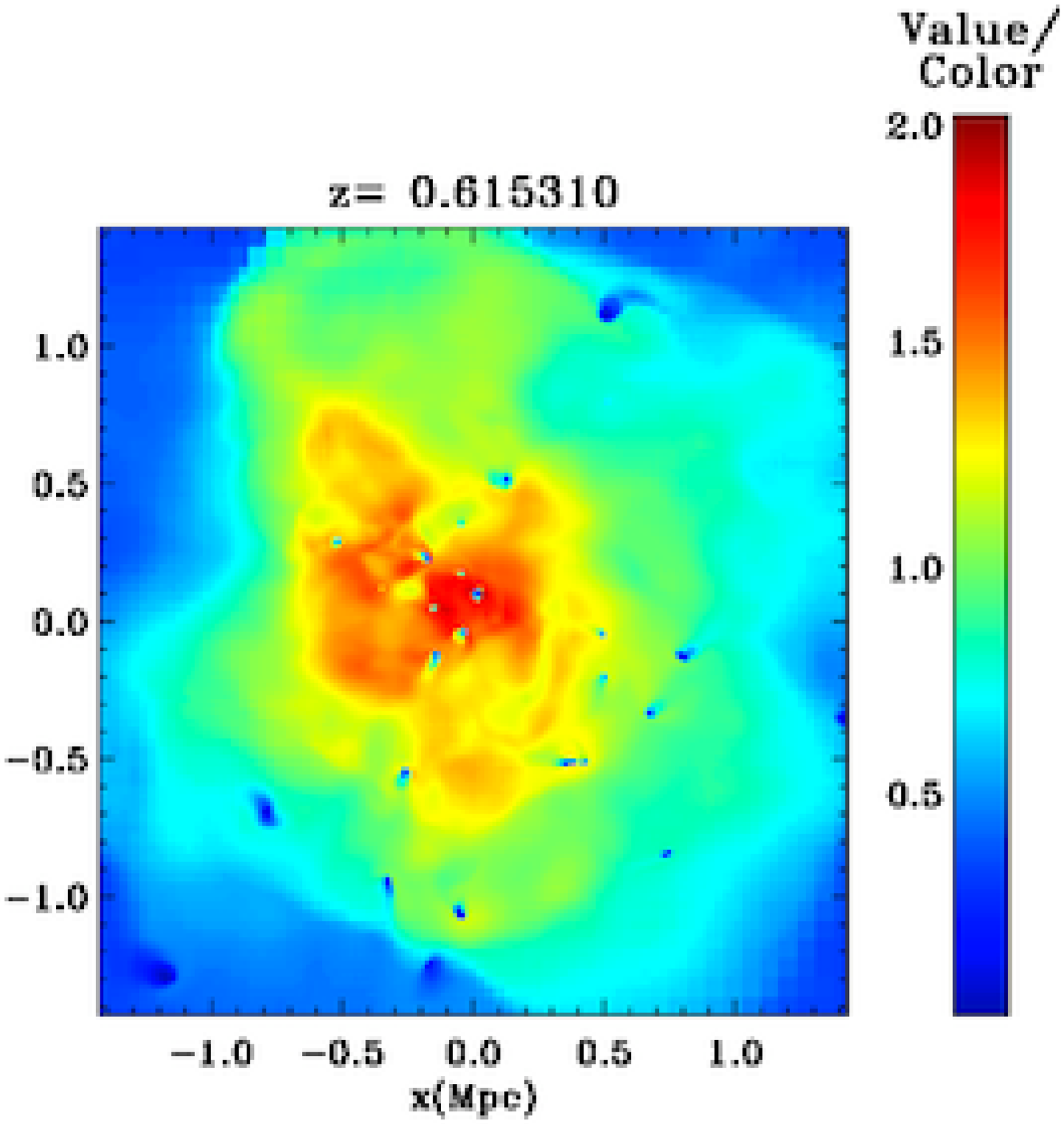}
    \includegraphics[width=0.45\textwidth]{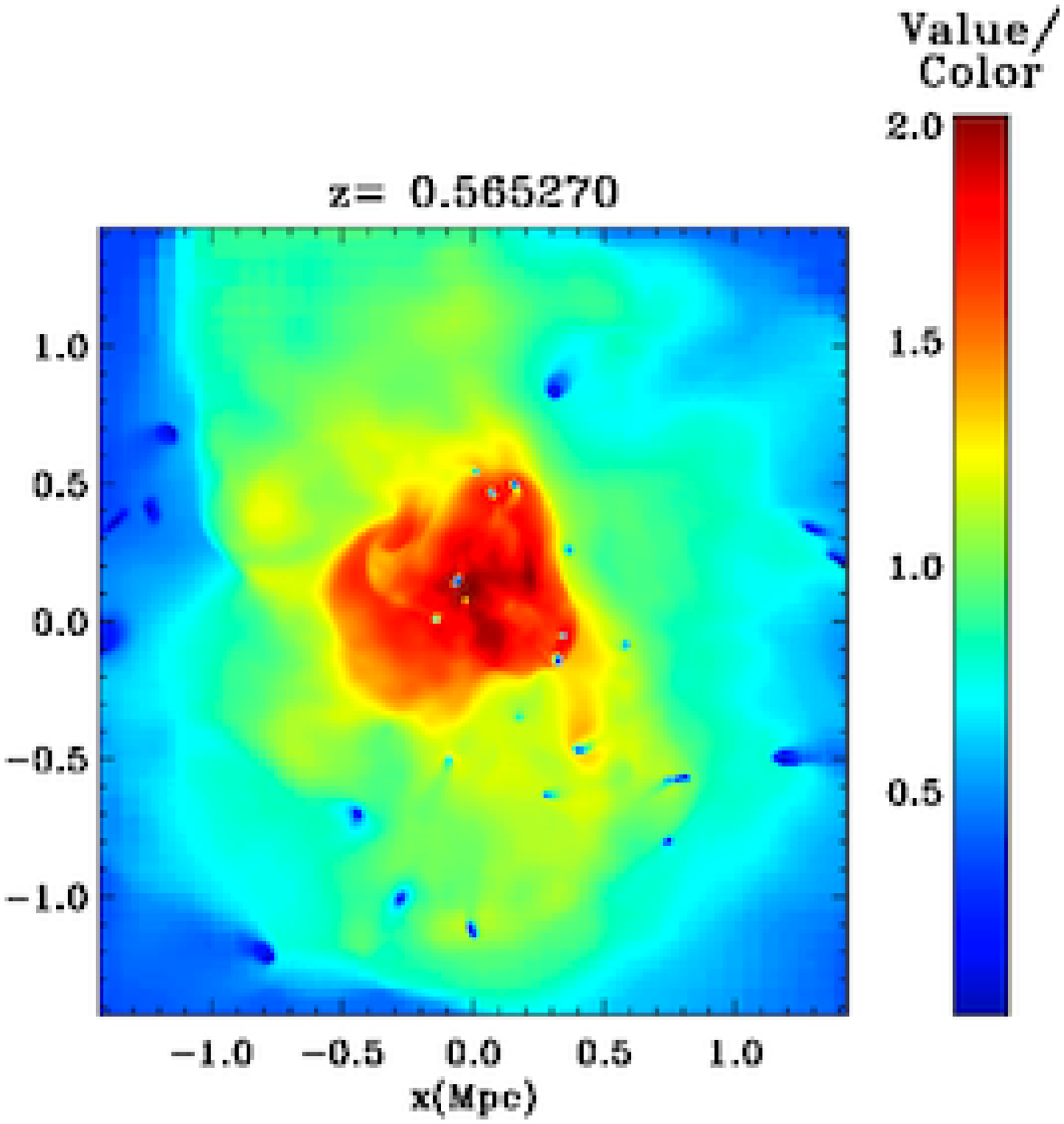}\\
    \includegraphics[width=0.45\textwidth]{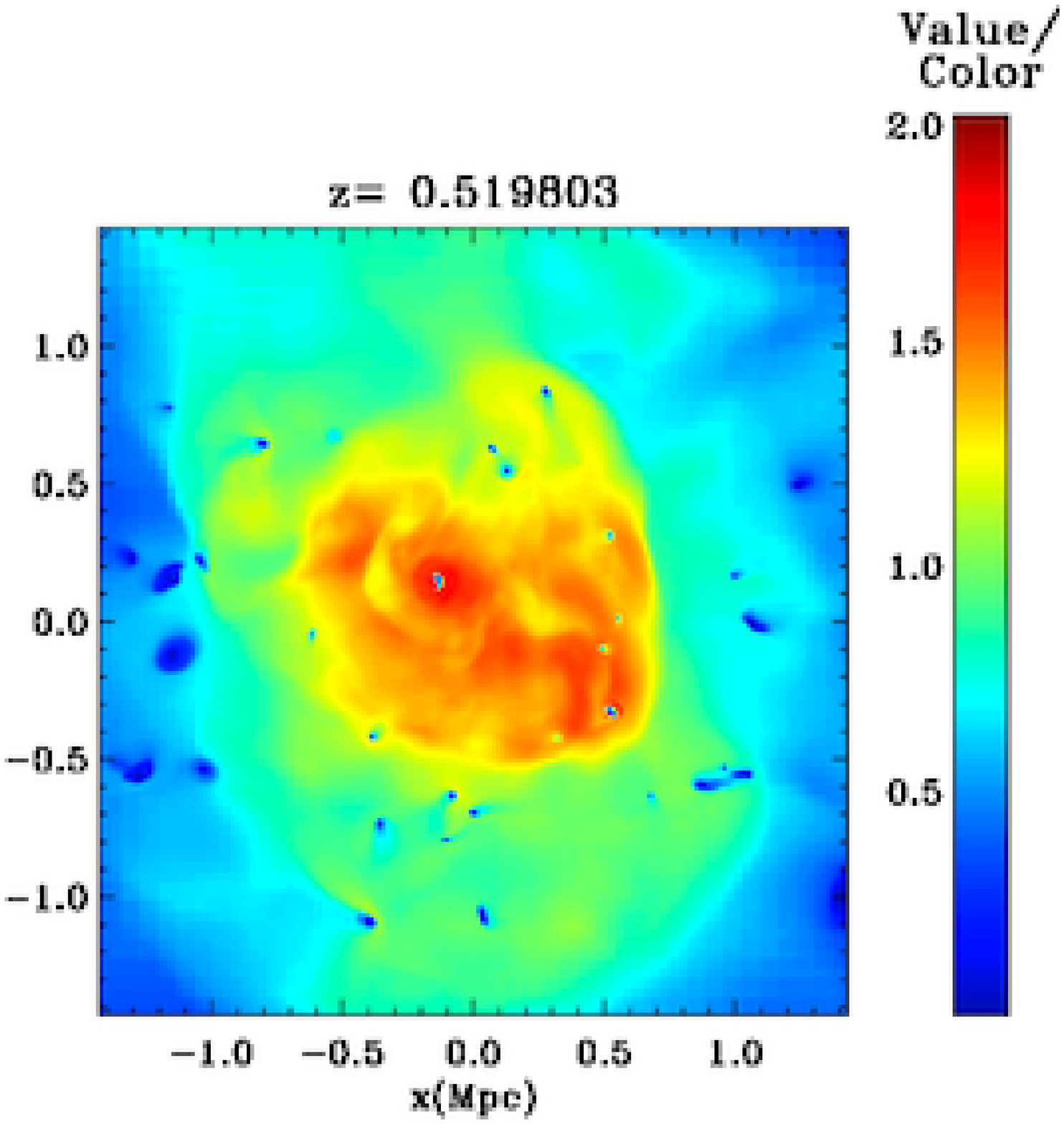}
    \includegraphics[width=0.45\textwidth]{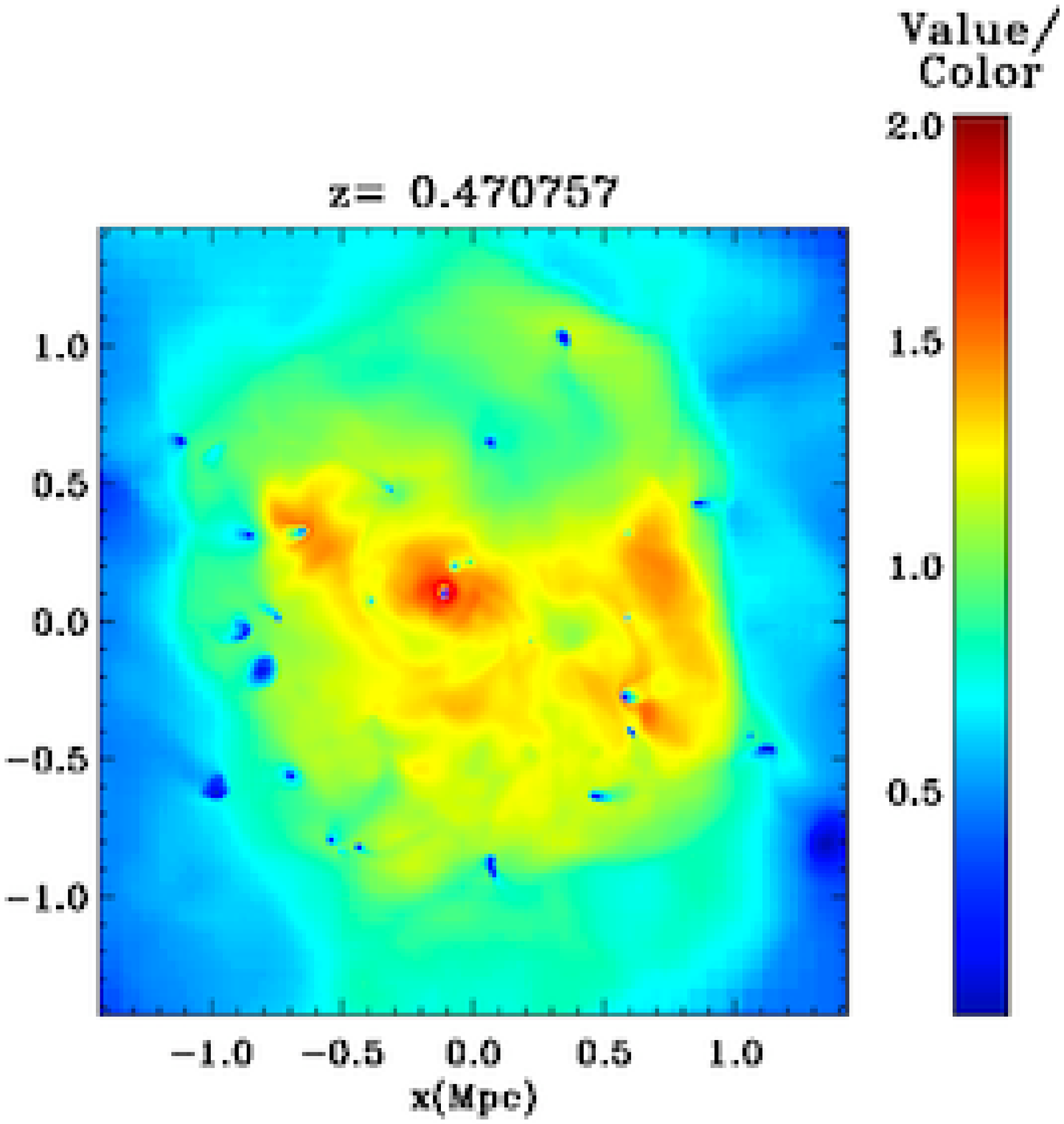}\\
  \caption{ Maps of the mass weighted temperature of the simulated cluster at four different redshifts. Units are in keV. The first redshift in the series corresponds to the strong SMBH outburst seen in Figure~\ref{fig:SmbhAcc} around $a \simeq 0.62$.}
  \label{fig:TempMaps}
\end{figure*}

In Figure~\ref{fig:AgnGrowth}, we show the accretion rate of the most massive SMBH as a function of expansion factor. This plot is relevant only for the "AGN" simulation. Also shown is the Eddington accretion rate, directly proportional to the SMBH mass. It appears quite clearly in this plot that the most massive SMBH grows discontinuously, during very short Eddington limited accretion events, or, at late time, by accreting other black holes, in good agreement with semi-analytical predictions from \cite{Malbon:2007p3846}. In the low resolution simulation, the final mass reaches $M_{\rm BH} \simeq 2.1\times 10^9$~M$_\odot$ after a last merger around $a \simeq 0.85$. In the high resolution simulation, the SMBH mass is twice as large, with a slightly different evolution, and reaches the final value of $M_{\rm BH} \simeq 4.2\times 10^9$~M$_\odot$. The SMBH mass in M87 has been estimated around $4 \times 10^9$~M$_\odot$ using dynamical constraints \citep{Macchetto:1997p1685, Gebhardt:2009p2, Gultekin:2009p475}, in very good agreement with our high resolution prediction. Note that M87 SMBH is close to the $M_{\rm BH}$--$\sigma$ relation \citep{Gultekin:2009p475}. Since the free parameter $\epsilon_c$ was calibrated to $\epsilon_c=0.15$ by \cite{Booth:2009p501} on the $M_{\rm BH}$--$\sigma$ relation, our AMR simulation appears to be consistent with their SPH results.

The SMBH activity, quite strong before $z=1$, declines slightly until the present epoch. In our simulation, this early activity is due to an early phase of frequent mergers feeding the SMBHs very efficiently. Strong and repeated outbursts of energy are launching strong shock waves in the IGM, rising the IGM entropy within the whole proto-cluster region. This early epoch can therefore be considered as representative of the pre-heating scenario advocated by several authors to explain structural properties of galaxy clusters \citep{Kaiser:1991p3983, Ponman:1999p4007, Babul:2002p3967, Dave:2008p3972}. On the other hand, at later epochs ($z<1$), when the cluster mass is finally assembled, AGN feedback prevent gas from overcooling and from accumulating in the core. This is well illustrated by the sequence of temperature maps shown in Figure~\ref{fig:TempMaps}, just after the strong AGN outburst occurring at $a \simeq 0.62$ (see Fig.~\ref{fig:AgnGrowth}). The first image show the mass-weighted projected temperature within the whole cluster, just at the time of the outburst. Slightly after (in the second frame), the whole cluster has been significantly heated, with buoyantly driven plumes of hot gas escaping the cluster core \citep{Chandran:2007p2323, Cattaneo:2007p420, Rasera:2008p2324, Sharma:2009p2272}. In the next frame, a strong shock, visible as a sharp temperature discontinuity, develops close to and beyond the virial radius. The last frame shows the cluster back to hydrostatic equilibrium, waiting for the next AGN outburst. Note that in the core region, where the density is high enough for X-ray detections, only buoyantly rising bubbles are seen. In our simulation, shock waves form only in the outer part of the cluster, with a Mach number of a few. Because of the rather low gas density, we believe that their detection is quite challenging, explaining why there is so little evidence of their presence \citep{Bourdin:2004p444}.

Although the feedback recipe doesn't explicitly account for it, the AGN energy deposition typically proceeds in two modes: a strong, energetic mode or "quasar mode", which in our case corresponds to the Eddington luminosity and reaches $5 \times 10^{46}$~erg~s$^{-1}$, and a quiescent mode or "radio mode", with a Bondi-Hoyle-limited luminosity of $5 \times 10^{41}$~erg~s$^{-1}$. If we now take into account the coupling efficiency parameter $\epsilon_c=0.15$ in our energy estimate, using, from Figure~\ref{fig:AgnGrowth}, $\dot{M}_{\rm acc} \simeq 10^{-4}$~M$_\odot$/yr into Equation~\ref{enerdump}, we obtain a total luminosity of $9 \times 10^{40}$~erg~s$^{-1}$ during radio mode. Although the X-ray luminosity of the active nucleus of M87 observed with {\it Chandra} is $L_{\rm X,0.5-7~keV} \simeq 7 \times 10^{40}$ erg~s$^{-1}$ \citep{DiMatteo:2003p4009}, quite close to our total luminosity in the radio mode, the mechanical power of the observed jet is significantly larger $P_{\rm jet} \simeq 3.3 \times 10^{43}$~erg~s$^{-1}$ \citep{Allen:2006p5821}. This is consistent with observational estimates of the Bondi accretion rate in M87, around  $\dot{M}_{\rm acc} \simeq 10^{-1}$~M$_\odot$/yr \citep{DiMatteo:2003p4009, Allen:2006p5821} and a coupling efficiency $\epsilon_c \simeq 0.2$ \citep{Allen:2006p5821}. This suggests that M87 SMBH lies in an intermediate state, between our radio and quasar modes. Nevertheless, the fact that our final black hole mass match well observational constraints of M87 is of great importance: it means that the {\it available} rest mass energy that the black hole can release into the forming cluster over its entire history is consistent with M87. This doesn't mean that the actual amount of energy deposited into the X-ray emitting gas {\it today} is correct. With these limitations in mind, we can now predict the global properties of our simulated Virgo cluster, in particular the mass distribution in stars, gas and dark matter.

\subsection{The mass distribution of stars and gas}

\begin{figure*}
    \includegraphics[width=0.4\textwidth]{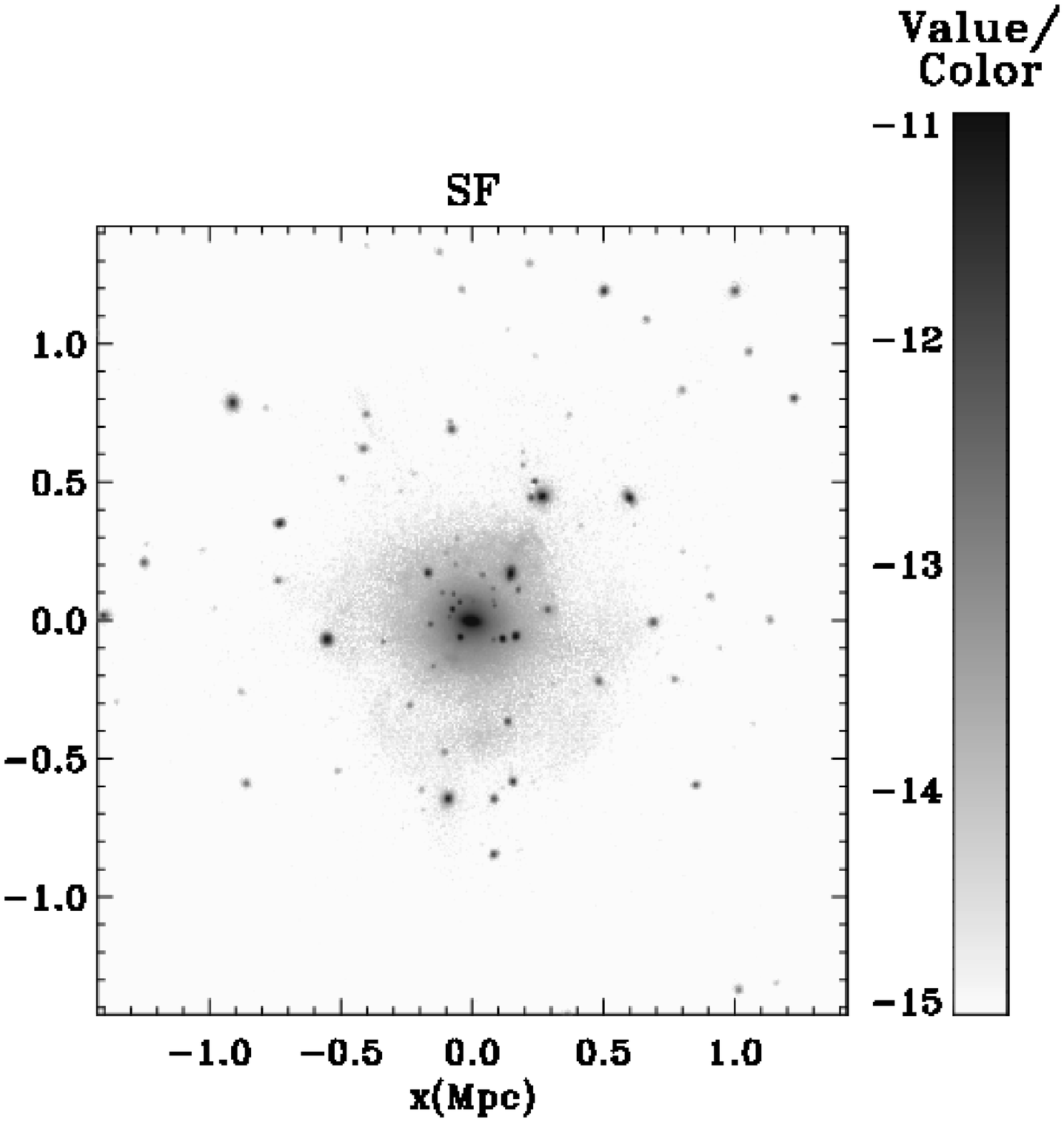}
    \includegraphics[width=0.4\textwidth]{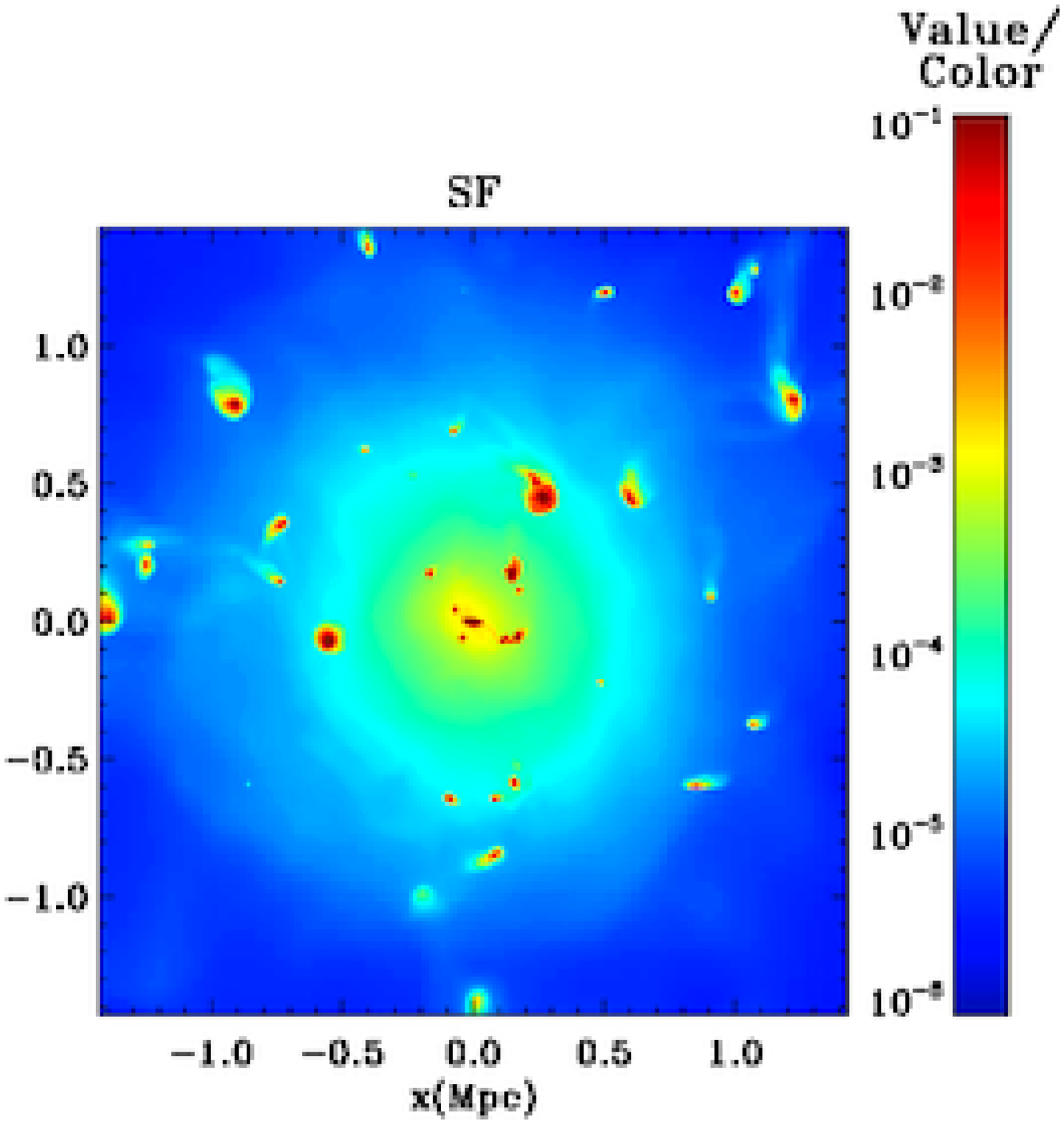}\\
    \includegraphics[width=0.4\textwidth]{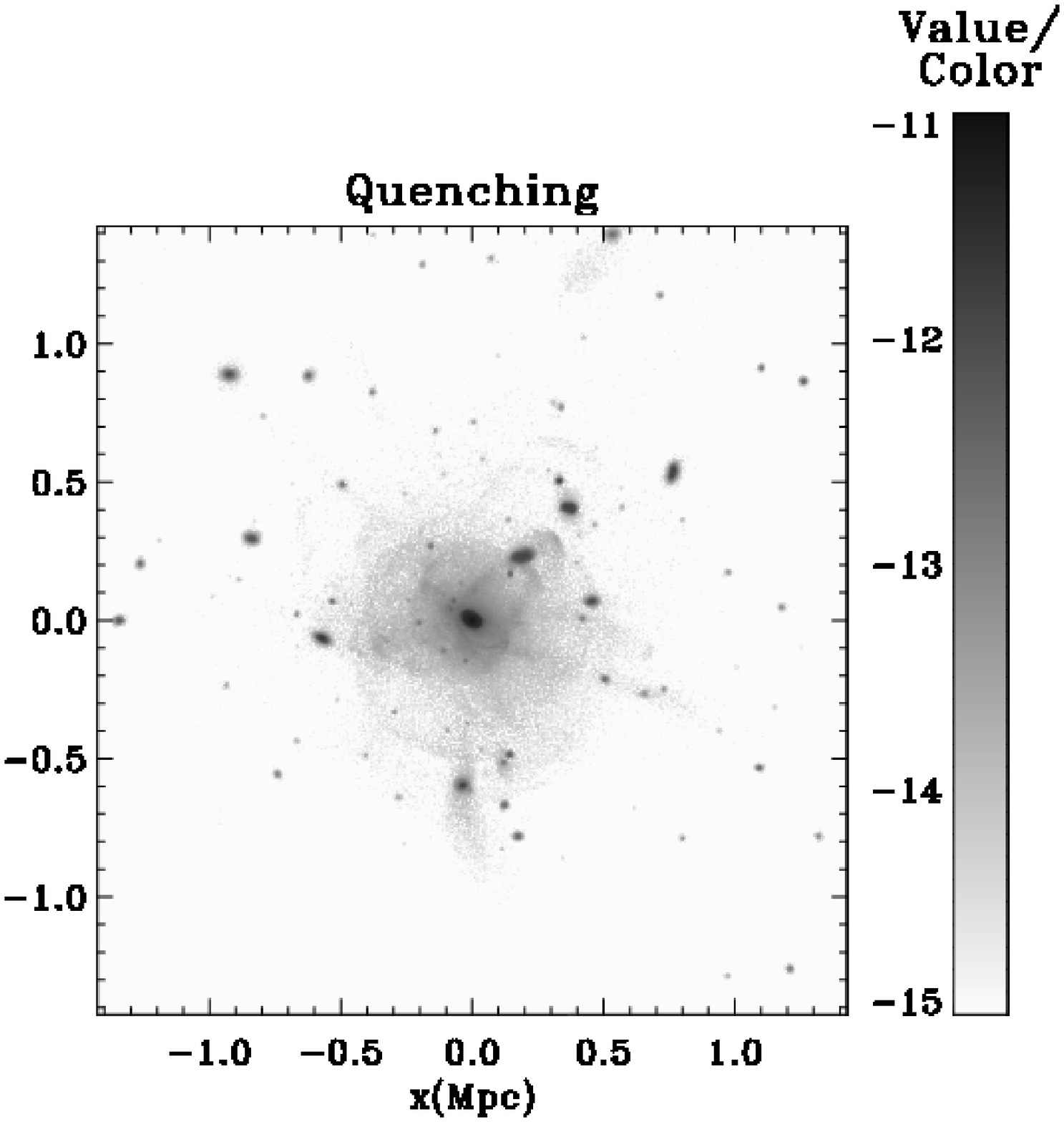}
    \includegraphics[width=0.4\textwidth]{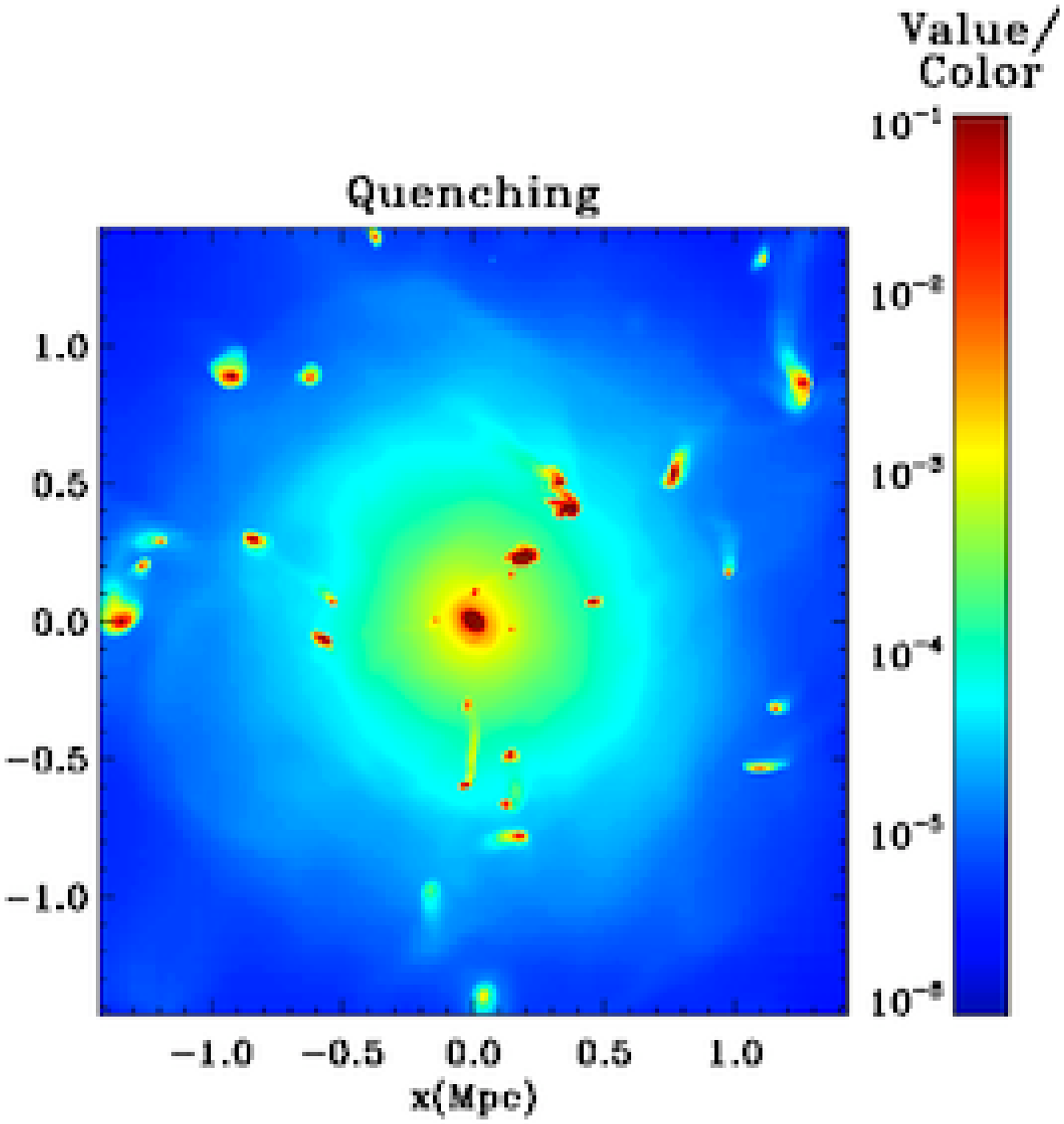}\\
    \includegraphics[width=0.4\textwidth]{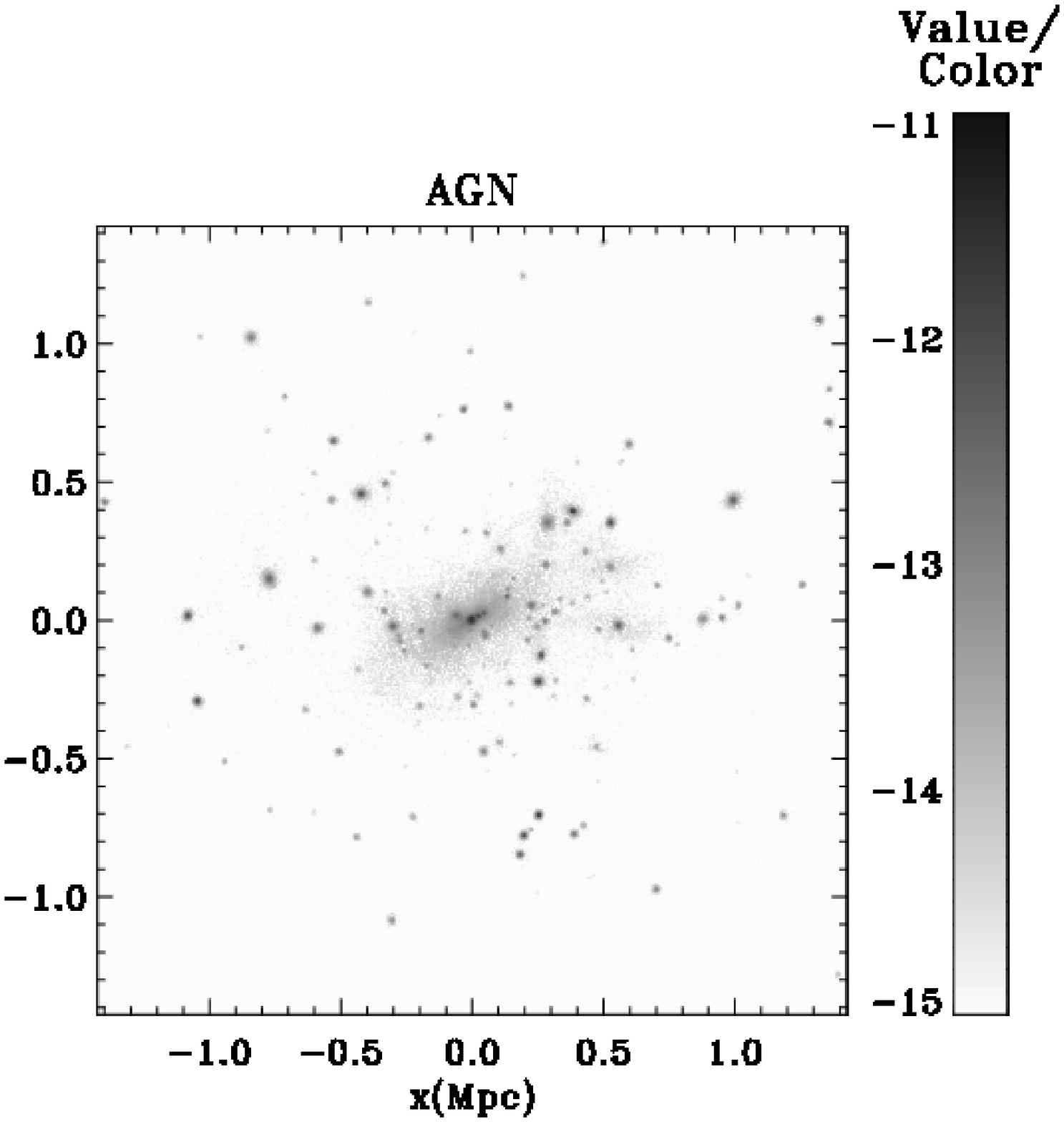}
    \includegraphics[width=0.4\textwidth]{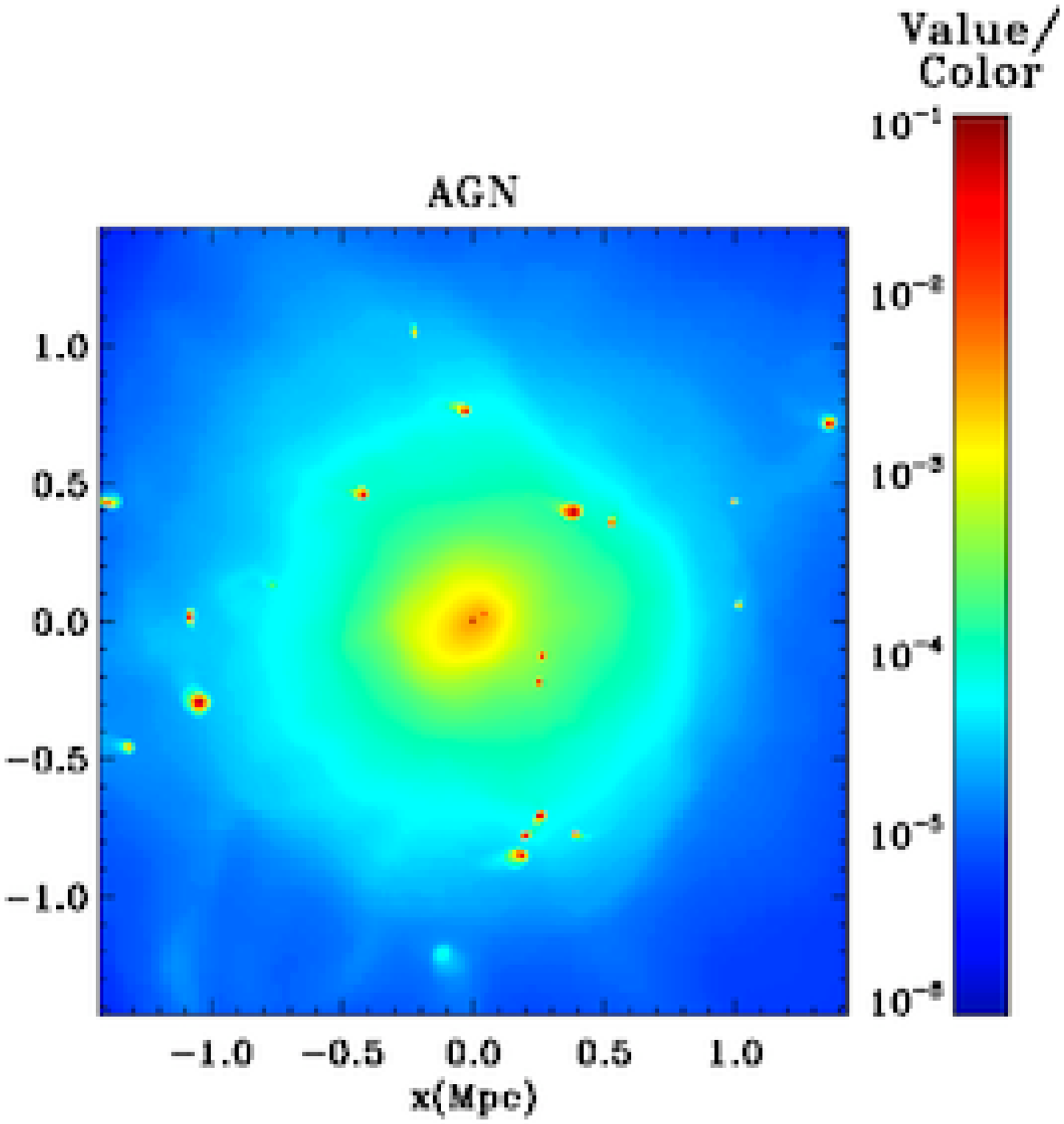}\\
  \caption{ \label{fig:StarGasMaps} Maps of the projected stellar luminosity (left panels, units are I band absolute magnitude) and gas mass weighted density (right panels, units are in H per cc) in the simulated cluster at $z=0$ for our three  models.}
\end{figure*}

\begin{figure}
  \begin{center}
    \includegraphics[width=0.45\textwidth]{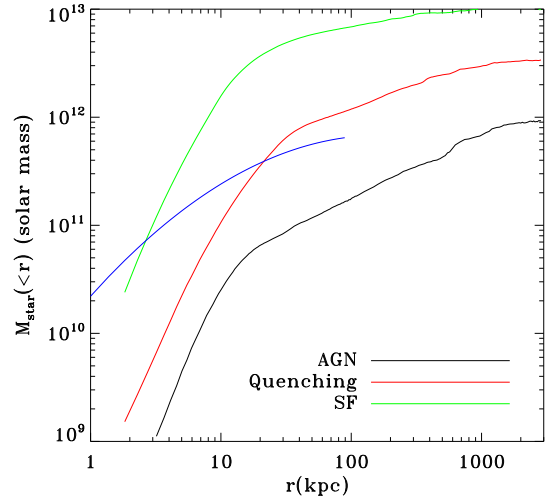}
  \end{center}
  \caption{ \label{fig:Star} Cumulative stellar mass profile for our simulated cluster at $z=0$. The blue line is the stellar mass profile of M87 from \citet{Gebhardt:2009p2}.}
\end{figure}

\begin{figure}
  \begin{center}
    \includegraphics[width=0.45\textwidth]{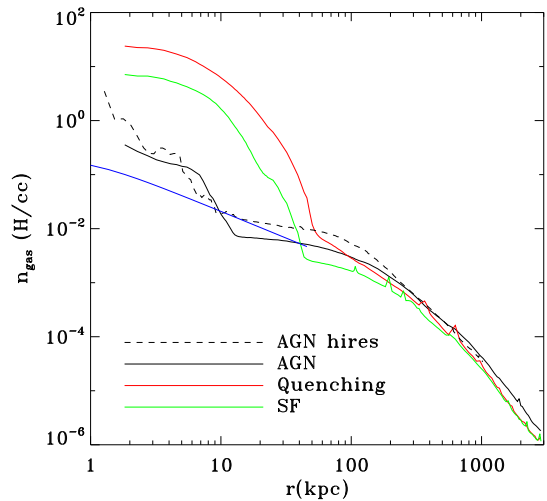}
  \end{center}
  \caption{ \label{fig:Gas} Gas density profile for our simulated cluster at $z=0$.  The blue line is the fit of the gas density using the X-ray emissivity profile of M87 from \citet{Churazov:2008p1}.}
\end{figure}

We have plotted in Figure~\ref{fig:StarGasMaps} the surface brightness of our simulated Virgo cluster in the SDSS i band.
One clearly sees many satellite galaxies orbiting around the BCG and a rich structure in the intracluster light component. In the SF case, the BCG appears as a very bright, gas rich 
and disc-like object. We have also plotted  in Figure~\ref{fig:Star} the stellar mass profile from this SF simulation: we see immediately that the BCG stellar mass (measured at 20 kpc from the centre) is a factor of 10 to large, when compared to the observational estimate of \cite{Gebhardt:2009p2}. This is the classical result of the over-cooling problem that occurs for cluster simulations with standard galaxy formation physics. Our second scenario, the Quenching run, partially alleviates  this problem. As can be seen on the surface brightness maps (Fig.~\ref{fig:StarGasMaps}), the BCG and the most massive satellites are now much dimmer. The total stellar mass within 100~kpc is in much better agreement with observations, although still slightly larger. The stellar mass profile, shown in Figure~\ref{fig:Star}, is however significantly different. When one looks now at the gas distribution in the cluster, it becomes quite obvious that this Quenching scenario is far from being a viable solution. We have plotted in Figure~\ref{fig:StarGasMaps} the mass-weighted, projected gas density. We see a massive gas clump in the cluster core for both the SF and the Quenching runs. For sake of comparison, we have plotted the simulated gas density profiles for our simulations and the best-fit $\beta$--model for M87 from  \cite{Churazov:2008p1}. The gas density is the SF run is a factor of 100 too large in the core of the cluster, and it is even worse for the Quenching scenario, by an additional factor of 2. We therefore conclude that for both SF and Quenching models, the simulated cluster suffers from a strong overcooling problem, with the build up of a dense, concentrated BCG, for which the stellar mass or the gas mass (or both) are in far in excess of those observed in M87.

We now turn to the analysis of our AGN model. The stellar surface brightness map is by far the dimmest of our 3 models: star formation has been dramatically reduced, even more than our Quenching scenario. The stellar mass profile is now below the observational constraints by a factor of 3 at 100 kpc (see Fig.~\ref{fig:Star}) and there is less apparent structure in the ICL component. AGN feedback has been quite successful in regulating star formation in the cluster. The gas distribution has also been profoundly affected by the SMBH model. First, no large, gas rich disc is visible in the projected gas density map: the overcooling problem have been efficiently removed. We see in Figure~\ref{fig:Gas} that the dense unrealistic gas core has disappeared. When compared to the best-fit $\beta$-model proposed by  \cite{Churazov:2008p1} for M87, the agreement is much better. Note that the cooling flow, although dramatically reduced, is still present in the AGN run, and it can be detected as the density enhancement within the central 10 kpc of the cluster. Interestingly, the knee in the gas density profile seen around 10 kpc in our model is also present in the data, although much weaker and at $\sim 5$ kpc from the center \citep[see Fig.~5 in their paper]{Churazov:2008p1}, suggesting the presence of a weak cooling flow in M87. Another important difference between the SF/Quenching runs and the AGN run can be seen at large radii in the gas distribution: the gas density in the AGN case is $30\%$ larger, showing that gas have been removed from the core and stored at large radii (beyond the virial radius) by strong shocks similar to the one shown in Figure~\ref{fig:TempMaps}. 

\begin{table}
\begin{center}
\begin{tabular}{|l|c|c|c|c|}
\hline
Run & $M^{\rm tot}_{200 \rm c}$ &  $f_{\rm bar}$ & $f_*$ & $f_{\rm gas}$ \\
\hline
SF & $1.2 \times10^{14}$ M$_\odot$ & $16\%$ & $8\%$ & $8\%$ \\
Quenching & $1.2 \times 10^{14}$ M$_\odot$ & $16\%$ & $3\%$ & $13\%$ \\
AGN & $1.1 \times 10^{14}$ M$_\odot$ & $13\%$ & 1\%& $12\%$ \\
\hline
\end{tabular}
\end{center}
\caption{Mass fractions inside the virial radius for our three different models. The universal baryon fraction we used in this paper is $15\%$.}
\label{tab:BaryonFraction}
\end{table}

\begin{figure}
  \begin{center}
    \includegraphics[width=0.45\textwidth]{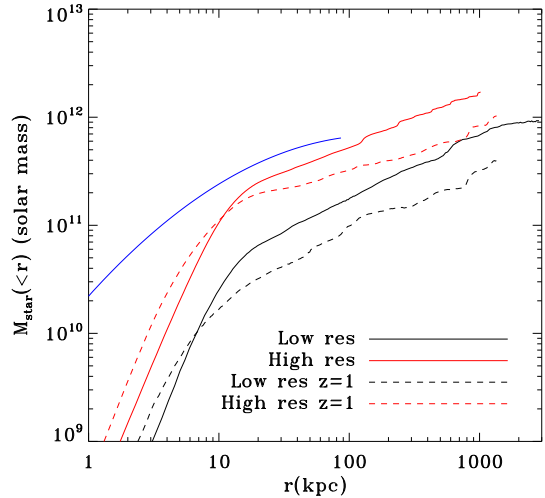}\\
  \end{center}
  \caption{ \label{fig:MstarResol} Effect of the mass resolution on the stellar mass profiles from our AGN model. The blue line is the stellar mass profile of M87 from \citet{Gebhardt:2009p2}.}
\end{figure}

\subsection{The effect of mass resolution}

Using our high resolution simulation with AGN feedback, we would like to estimate the effect of numerical resolution on our results. We see in Figure~\ref{fig:MstarResol} the stellar mass profiles for the low and the high resolution runs at the final redshift. Contrary to what is often claimed in the literature, we do see a strong effect of mass (and spatial) resolution in the stellar mass distribution. The effect is stronger for the BCG close to the center (the stellar mass has increased by a factor of 4 at 20 kpc) than for the cluster as a whole (the total stellar mass within the virial radius has increased only by a factor of 2). The effect of mass resolution is smaller for the other components (gas and dark matter). 

The strong effect of mass resolution on the star formation history of the simulated halo can be interpreted easily by comparing the minimum resolved halo mass (optimistically set to 100 dark matter particles) to the minimum mass for star forming halos based on atomic cooling arguments \citep{Gnedin:2000p1555, Rasera:2006p392, Hoeft:2006p1565}. This minimum mass (also referred to as the Filtering Mass) starts around $10^7$ M$_\odot$ before reionization and then rises steadily as $(1+z)^{3/2}$ from redshift 6-7 to the final epoch.
Resolving this minimum mass before reionization will require a dark matter particle mass below $10^5$~M$_\odot$, a rather strong requirement for cluster--scale cosmological simulation. A more flexible criterion based on resolving the majority ($\sim 80\%$) of star forming halos gives a less stringent limit around $M_{\rm min} \simeq 10^{8}$~M$_\odot$ \citep{Iliev:2007p1603}. Nevertheless, our low resolution run falls short of the corresponding required dark matter particle mass by 65, while our high resolution run is "only" a factor of 8 above the limit. As explained in \cite{DeLucia:2007p1656}, BCG are "fundamentally hierarchical" objects, that formed their stars very early ($80\%$ before $z=3$) and assembled late (after $z=0.5$ in average). This effect is directly related to the suppression of cooling flows and the associated star formation by AGN feedback. This early star formation occurs in rather small mass halos \citep{DeLucia:2007p1656} in which star formation proceeds through accretion of diffuse gas in cold streams \citep{Dekel:2009p359}.  Although BCGs are quite massive objects, it is of great importance to resolve properly the earliest epoch of star formation, in order to account for all the stellar mass in these objects. \cite{Puchwein:2010p763} have reported a similar effect in their SPH simulation, although they mentioned a significantly smaller effect ($\sim$ 20\%) with a different, may be more robust feedback model.  

Another interpretation to the rather large resolution effect we see in our simulation is an evolution in the efficiency of AGN feedback. The AGN thermal energy deposition is performed within a sphere of 4 cells radius. The high resolution run will therefore deposit the energy deeper into the halo potential well. This might result in a reduced overall efficiency. Using the same model than in this paper,  \cite{Booth:2009p501} have also reported a rather strong dependence of the computed star formation rate on mass resolution (see their Fig.~6b). Another solution we would like to explore in the future is to recalibrate the AGN feedback model parameters as a function of mass resolution, in order to overcome these limitations.

For both the low and the high resolution simulations, we see in Figure~\ref{fig:MstarResol} that the stellar mass profile has evolved only slightly between $z=1$ and $z=0$. Inside the BCG, we see stars expanding slightly, while the stellar halo grows in mass more substantially, by almost a factor of 2. This evolution is in good qualitative agreement with the group--scale simulation reported by \cite{Feldmann:2010p1516}. Using our highest resolution simulation, we observe that the agreement with M87 stellar distribution is very good between 10 and 100 kpc. It however still deviates quite strongly with observations within the inner 10~kpc. We therefore conclude that our model with AGN feedback seems to converge {\it from below} to the correct stellar mass distribution. It is worth stressing that the same analysis can be made for our Quenching run, and that its converged stellar mass ends up being significantly above the observational limit. From this, we conclude that AGN feedback is necessary, not only to regulate star formation inside massive galaxies, but also to destroy and remove the gas supply in satellite galaxies.  

\subsection{The distribution of dark matter}

\begin{figure*}
    \includegraphics[width=0.45\textwidth]{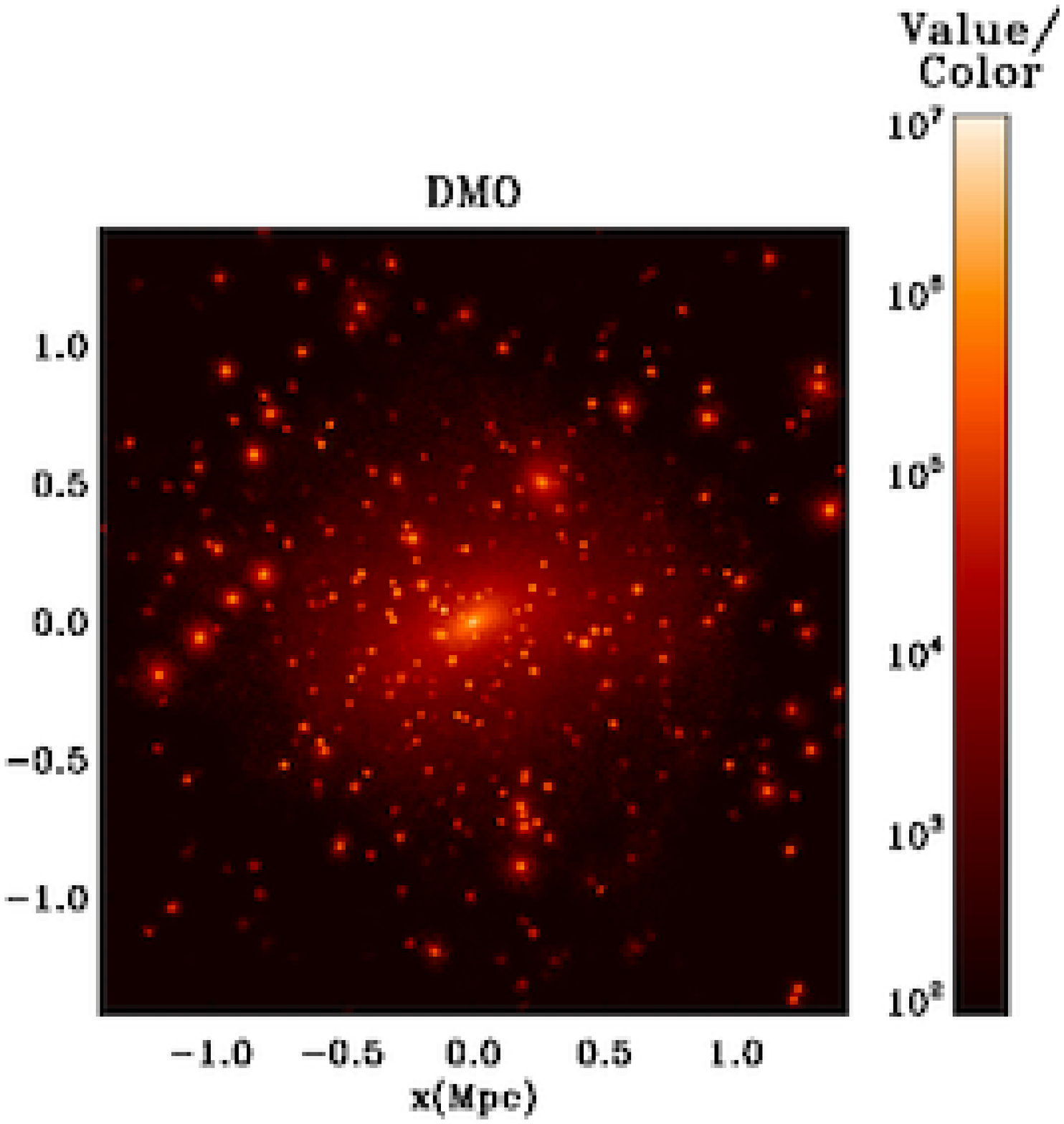}
    \includegraphics[width=0.45\textwidth]{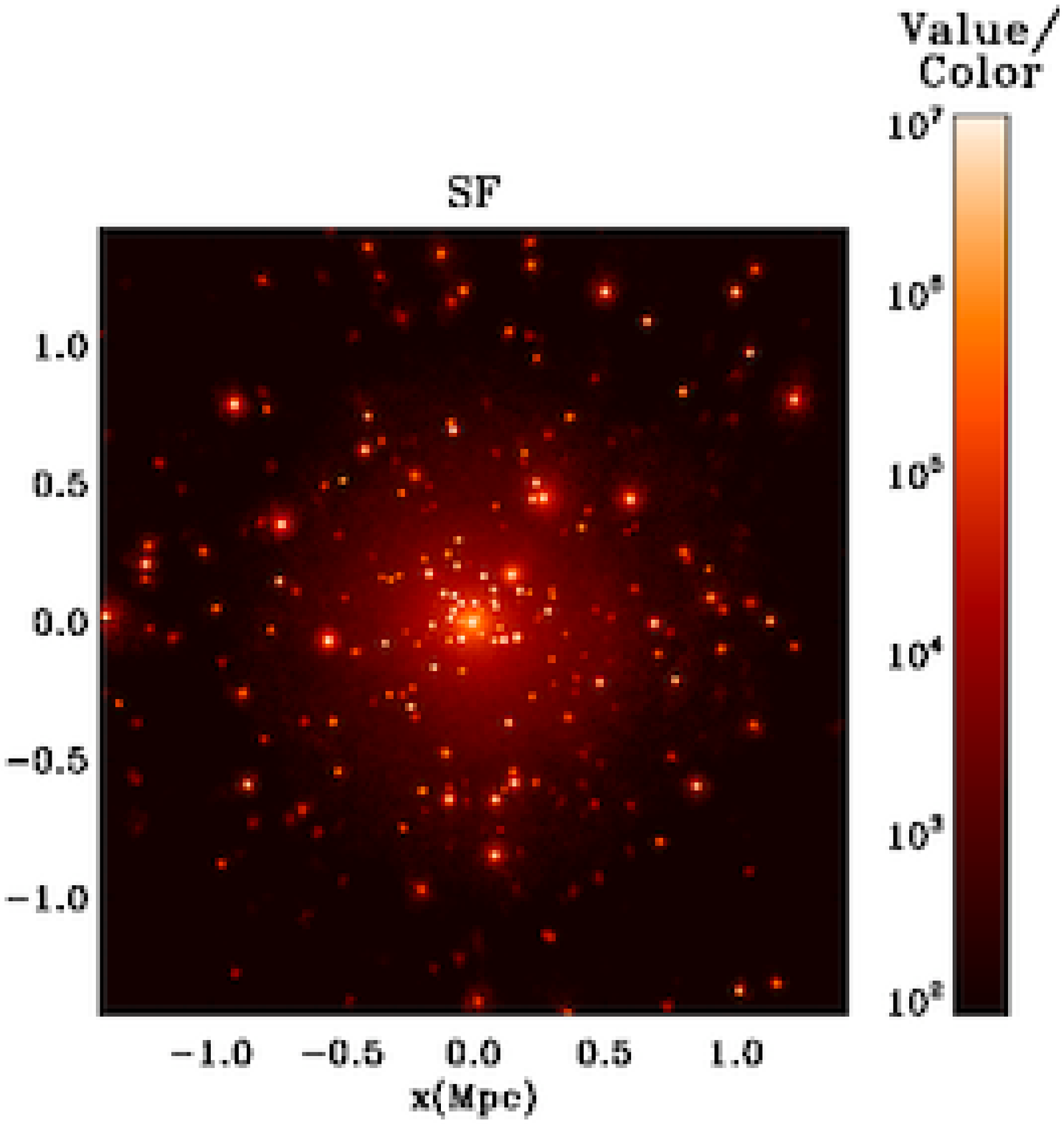}\\
    \includegraphics[width=0.45\textwidth]{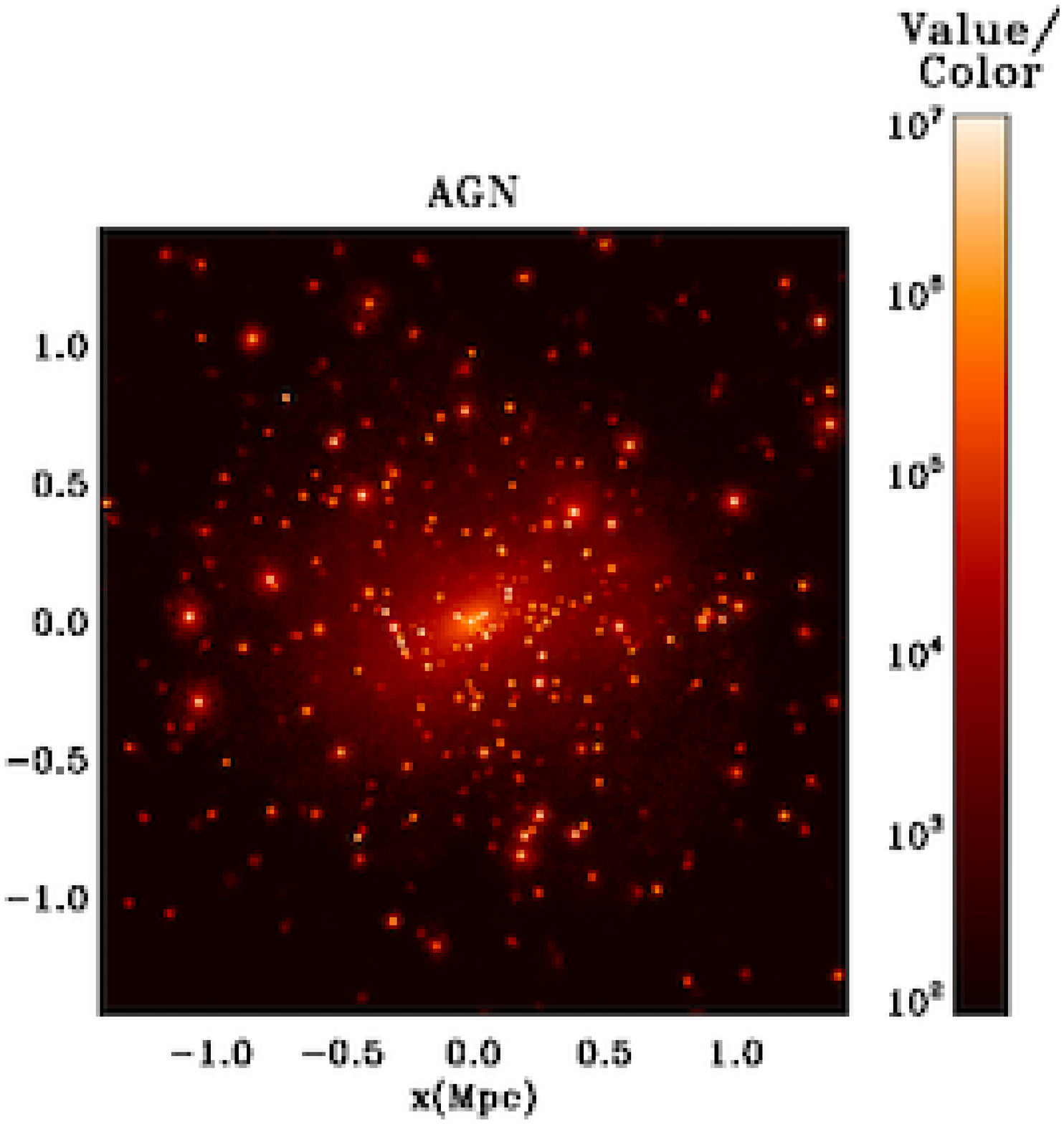}
    \includegraphics[width=0.45\textwidth]{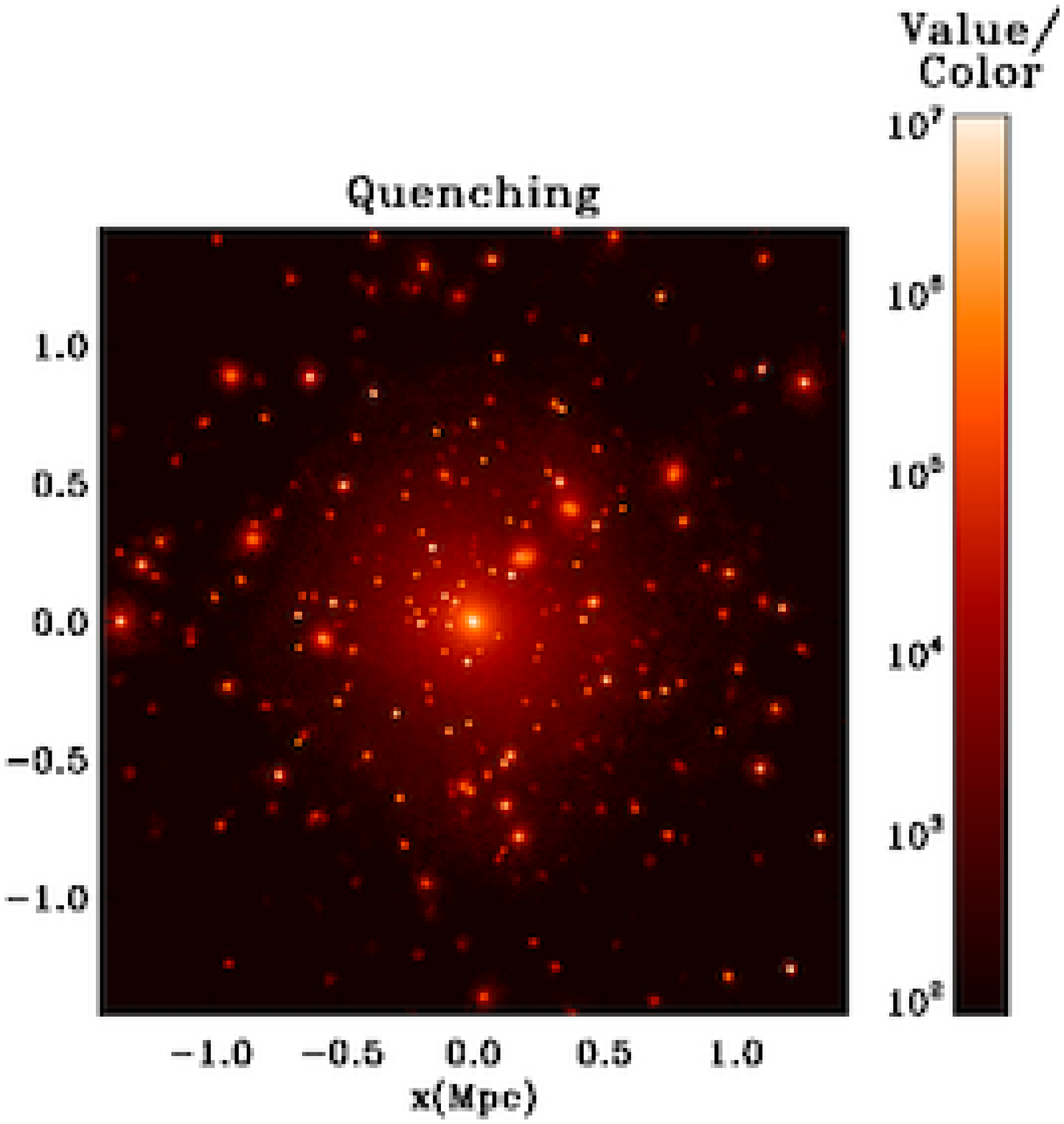}\\
  \caption{ \label{fig:CdmMaps} Maps of the dark matter overdensity distribution in the simulated cluster at $z=0$ for our 3 models. Also shown for comparison is the dark matter density map we have obtained in a dark matter only (DMO) simulation of the same cluster.}
\end{figure*}

\begin{figure}
  \begin{center}
    \includegraphics[width=0.45\textwidth]{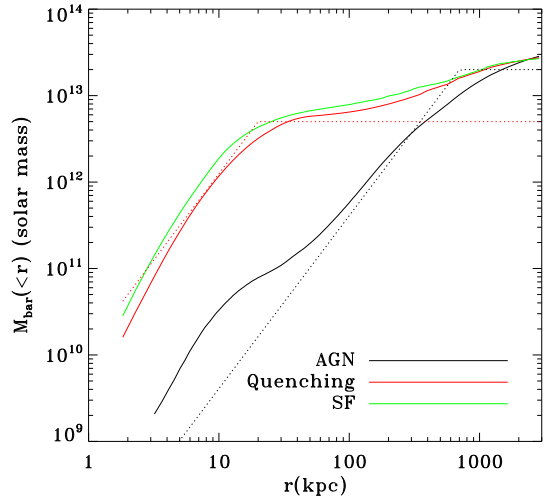}\\
  \end{center}
  \caption{ \label{fig:Mbar} Cumulative total baryonic mass profile (gas + stars) in the simulated cluster at $z=0$. The dotted lines are fits to the measured profiles for our simple model of adiabatic contraction of the dark halo.}
\end{figure}

\begin{figure}
  \begin{center}
    \includegraphics[width=0.45\textwidth]{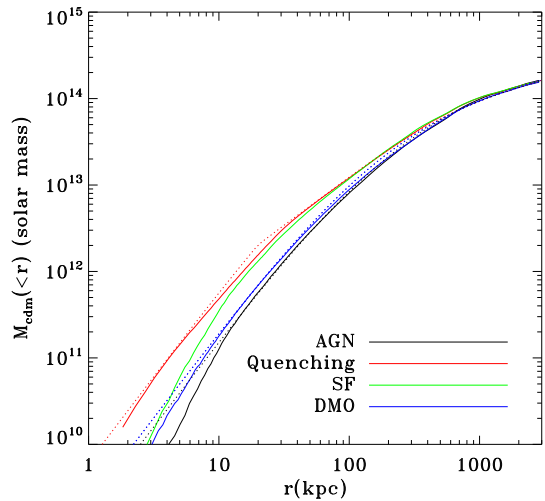}\\
  \end{center}
  \caption{ \label{fig:CdmMass} Cumulative dark matter mass measured in our three models at $z=0$. Also shown in blue is the profile we obtained in the dark matter only (DMO) simulation. In each case, the dotted line corresponds to our analytical model of adiabatic contraction. Note that in the AGN feedback case, we see a slight adiabatic expansion of the dark halo.}
\end{figure}

One important consequence of the overcooling problem is to modify significantly the properties of the dark matter halo. We have plotted in Figure~\ref{fig:CdmMaps} the projected dark matter density for our 3 models, and for the corresponding pure dark matter simulation. We immediately see that the dark halos in the overcooled runs (SF and Quenching) are denser and rounder than the pure dark matter case. With AGN feedback, we basically recover the halo shape and distribution of the pure dark matter case. The effect of gas cooling on the dark halo has been interpreted in term of {\it adiabatic contraction} of the particle orbits \citep{Blumenthal:1986p732, Gnedin:2004p569}, meaning that individual orbits are compressed inward, while conserving the adiabatic invariant ${\cal I}=rM(<r)$. The global shape change has been also interpreted by \cite{Debattista:2008p634} as a transition from boxy orbits to more circular ones. These effects have been studied quite extensively at galactic scales, where they are probably more relevant \citep{Abadi:2009p531, Pedrosa:2009p598}. Although we are dealing with a much larger object, we recover very similar properties for our dark halo, because of overcooling. We have plotted in Figure~\ref{fig:CdmMass} the cumulative dark matter mass profiles for our three runs, plus the Dark Matter Only (DMO) simulation. The DMO profile has been multiplied by $85\%$ to allow a direct comparison with the baryonic runs: it is fitted with an accuracy better than 5\% down to 10~kpc by a NFW profile with a concentration parameter $c=7.5$. The fit is shown as the blue dotted line on the same figure. The dark matter profiles for the SF and Quenching runs are very similar, except in the very centre of the cluster. They all appear significantly adiabatically contracted. Interestingly enough, the AGN case appears slightly expanded, when compared to the DMO simulation. We will now use AC theory to explain these trends.

\cite{Gnedin:2004p569} have revisited the original paper of \cite{Blumenthal:1986p732} on AC theory, stressing that the original assumption of purely circular orbits was leading to an overestimate of the baryon-induced dark halo contraction. They presented a numerical  implementation for their modified AC theory in which the particle orbit distribution was allowed some radial components and predicted the contracted dark matter distribution, given the initial dark matter profile and the final baryonic distribution. We present in the Appendix a simple analytical model to account for the adiabatic contraction of the dark halo, based on \cite{Gnedin:2004p569} theory. 

In Figure~\ref{fig:Mbar} we have plotted the total baryonic mass $M_{\rm bar}$ as a function of the final radius for our 3 different models. Although the actual distributions are different from our simple model, we have fitted them using a constant surface density, truncated  disc of mass $m_d$ and size $r_d$. The fits are very similar in the SF and Quenching case, so we have used only one model with $m_d \simeq 5\times 10^{12}$~M$_\odot$ and $r_d \simeq 20$~kpc. These values, consistent with the mass of our simulated BCG (including the concentrated gas component in the Quenching case) are much larger than any observed BCG in halos of similar masses, and are again a manifestation of the overcooling problem. In this case of strong baryonic concentration, our analytical model with $r_d \ll r_s$ applies (see the Appendix): we have plotted the corresponding AC theory prediction in  Figure~\ref{fig:CdmMass} as the dotted line, showing convincingly that the \cite{Gnedin:2004p569}  model, in our simplified formulation, works quite well: the fit is better than 20\% down to 10~kpc in radius. Within 10~kpc, our fit is not as good. We believe that since we are entering the scale of both the disc and the dark matter inner region ($r<r_s$), the adiabatic contraction model doesn't apply anymore. When the halo first forms and virializes at high redshift, the baryon fraction in star and gas during violent relaxation is likely to play an important role, explaining why the dark matter profiles in the SF and Quenching runs differ below 10~kpc, although the total baryon mass profiles are very similar at redshift zero.

The AGN feedback model gives us a much more extended baryonic mass profile. We have fitted it using the constant surface density truncated disc model with parameters $m_d \simeq2\times 10^{13}$~M$_\odot$ and $r_d \simeq 700$~kpc. We can see in Figure~\ref{fig:Mbar} that our fit is far from being perfect (especially within the BCG), but it captures roughly the total baryon distribution, which is now mostly dominated by the hot extended gaseous halo. The values for $m_d$ and $r_d$ are now closer to the total mass and total size of the hot halo than those of the BCG. In this case, the AC prediction cannot be worked out analytically. We have therefore solved for the real root of the third order polynomial defined by Equation~\ref{equ:Abadi} and plotted the result in Figure~\ref{fig:CdmMass}. Again, the prediction from \cite{Gnedin:2004p569} theory is very close (within 20\% above 10~kpc) to the measured dark matter profile. In the AGN feedback case, we see that the dark matter has expanded slightly, when compared to the pure dark matter case, and that this "adiabatic expansion" appears to be  well captured by the same adiabatic invariant for the orbits of dark matter particle. Note that this expansion is very small, as it affects the dark matter mass distribution by less than a few percent. We have also checked that using the complete numerical solution of  \cite{Gnedin:2004p569} in conjunction with the simulated baryon mass profile (instead of our simple disc model) does not affect our conclusions.

\begin{figure}
  \begin{center}
    \includegraphics[width=0.45\textwidth]{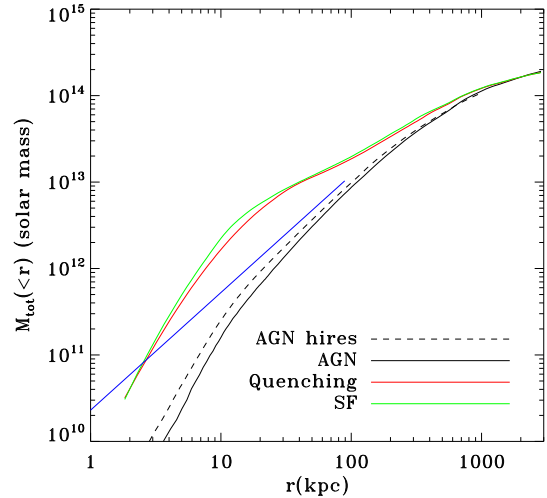}\\
  \end{center}
  \caption{ \label{fig:MassTot} Cumulated total mass profile (baryons + dark matter) in the simulated cluster at $z=0$. The blue line in the is the mass profile deduced from stellar kinematics and X-ray data \citep{Gebhardt:2009p2}.}
\end{figure}

\begin{figure}
  \begin{center}
    \includegraphics[width=0.45\textwidth]{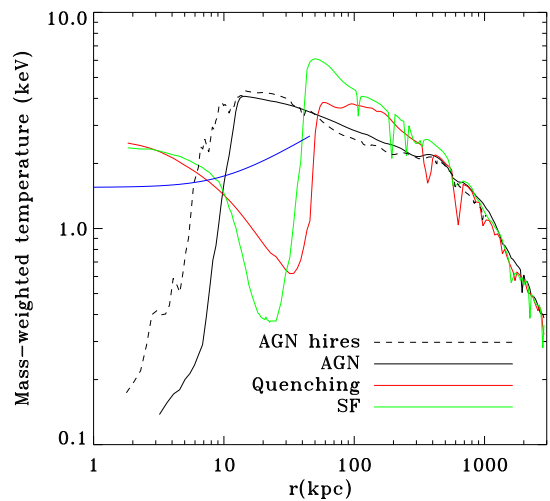}\\
  \end{center}
  \caption{ \label{fig:Temp} Mass-weighted gas temperature profile in the simulated cluster at $z=0$. The blue line in the is the temperature profile inferred from X-ray data in Virgo \citep{Churazov:2008p1}.}
\end{figure}

A good diagnostic of how much concentrated the simulated halo should be to match the observations is to compare the total mass profile with the one derived from dynamical arguments using M87 optical and X-ray data (see Fig.~\ref{fig:MassTot}).  \cite{Gebhardt:2009p2} have derived a complex mass model for M87, including the presence of a central SMBH, stellar BCG and dark matter halo. Their mass model is compared to our total mass profiles from our three different models in Figure~\ref{fig:MassTot}. The SF and Quenching models are too much concentrated and overestimate the total mass by a factor of 2 to 4 in the inner 100 kpc. The AGN model is in much better agreement, although slightly below, for both the low and high resolution runs.

Another diagnostic on the total mass distribution is the temperature profile of the hot, X-ray emitting gas. We have plotted in Figure~\ref{fig:Temp} the mass-weighted temperature profile for our various simulations. The temperature at 100~kpc is 3~keV in the AGN case, in good agreement with the observed value of 2.8$\pm$0.2~keV \citep{Churazov:2008p1}. Without AGN feedback, we obtain a larger temperature (around 6~keV) because of the higher mass concentration. If one looks at the temperature profile, we do not get a good match with X-ray data, even including AGN feedback. We are slightly hotter than the observations between 10 and 100~kpc, and within 10~kpc, we see a sharp drop of temperature, due to cooling gas feeding the central galaxy and sinking towards the central AGN. We have checked that hydrostatic equilibrium is satisfied down to 10~kpc. Matching X-ray temperature profiles will probably requires additional physics, such as cosmic rays propagation and magnetic fields, and/or a more realistic AGN feedback scheme.
 
\subsection{The baryon fraction}

\begin{figure}
  \begin{center}
    \includegraphics[width=0.45\textwidth]{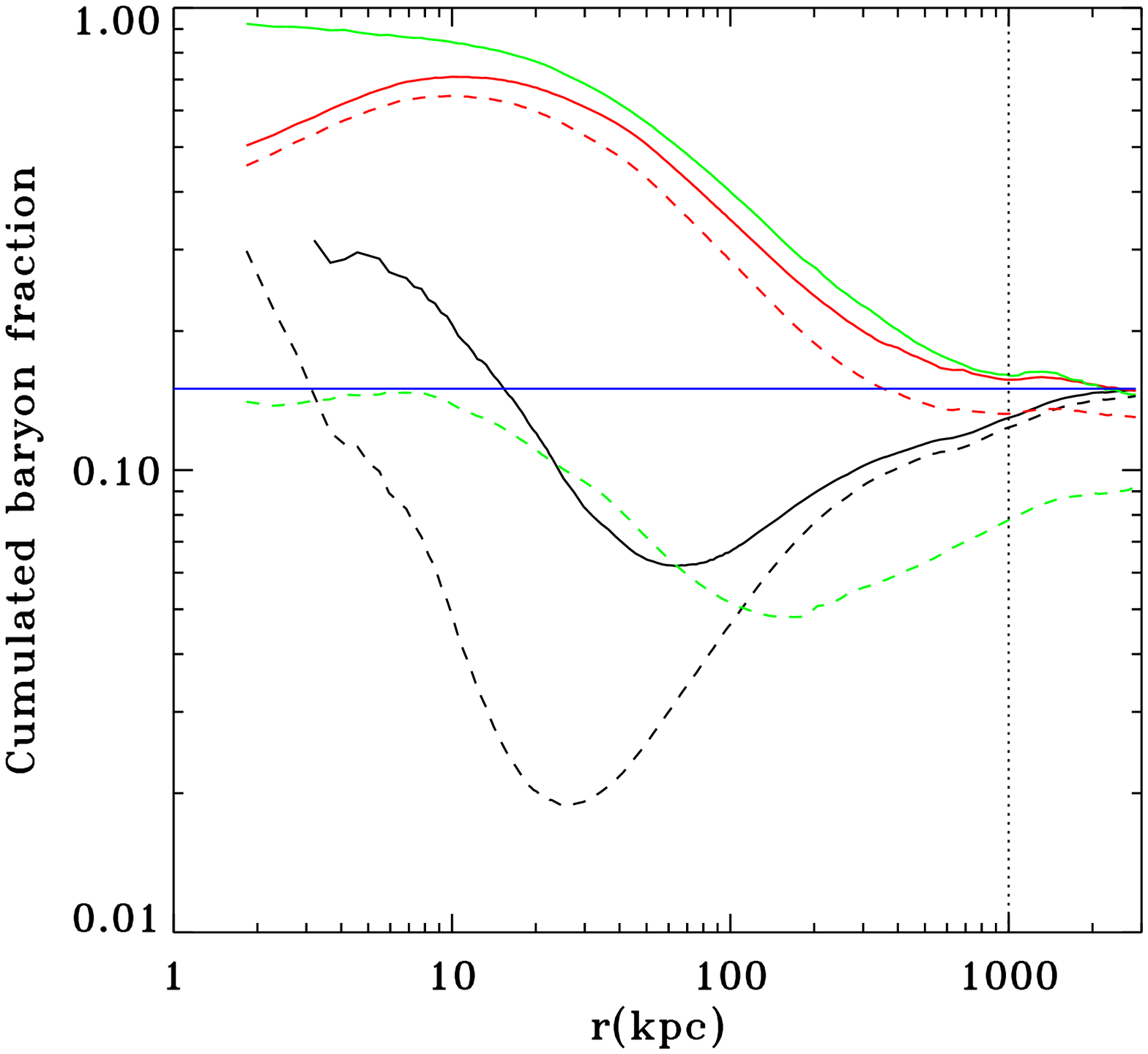}\\
    \vspace{2pt}
    \includegraphics[width=0.45\textwidth]{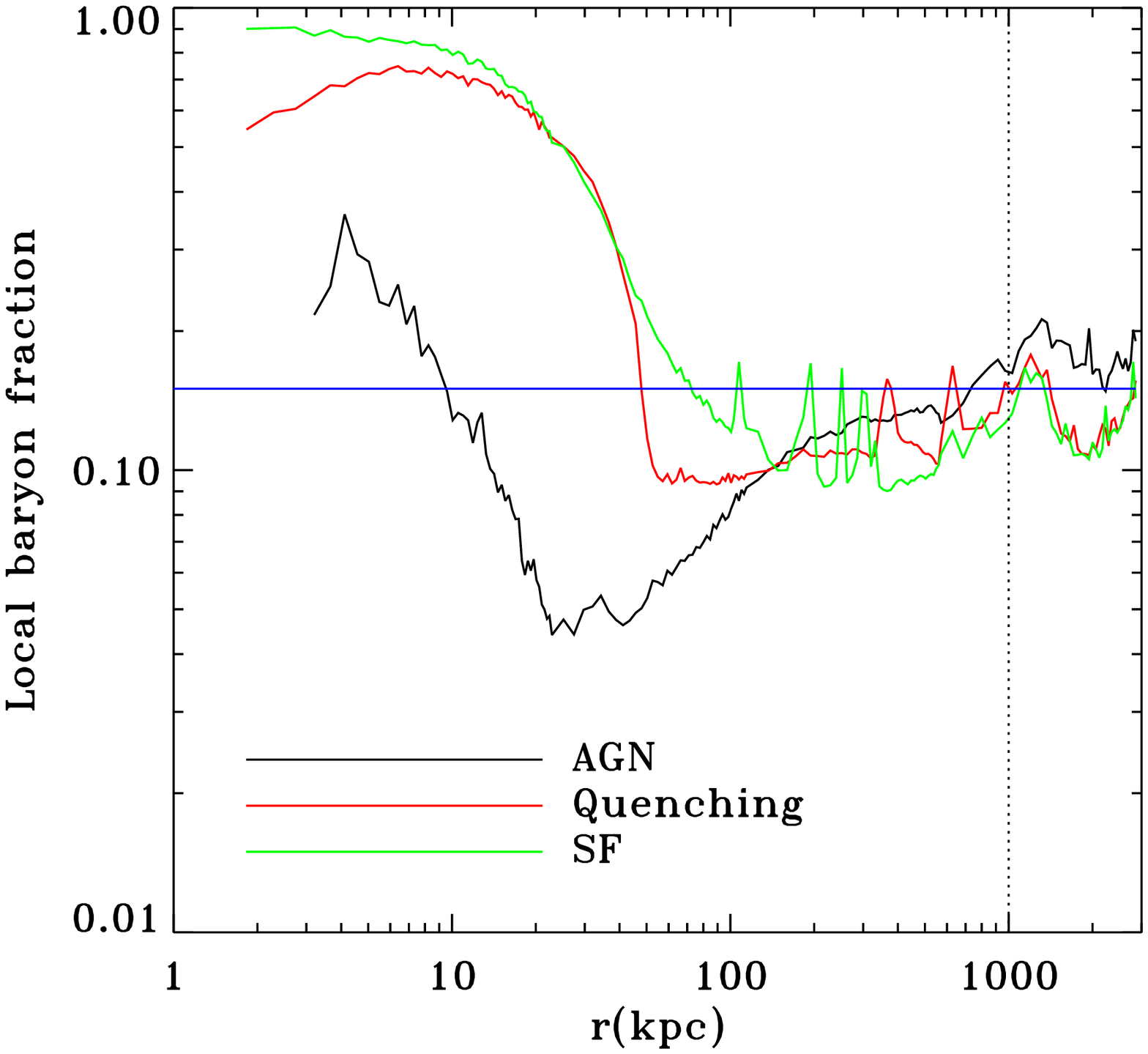}\\
  \end{center}
  \caption{ \label{fig:BaryonFraction} Cumulative (top panel) and local (bottom panel) baryon fractions at redshift $z=0$ in the simulated cluster. The universal baryon fraction is shown as the blue horizontal line, while the virial radius is indicated by the dashed vertical line. Each model is labelled using the standard color scheme. In the AGN feedback case, we observe a slight deficit ($\sim$10\%) of baryon within the virial radius. Using the local baryon fraction, we see that these missing baryons are located in between $R_{vir}$ and $2R_{vir}$. In the upper panel, the cumulated gas fraction is also shown as a dashed line.}
\end{figure}

The most important consequence of the disappearance of the overcooling problem is that we expect the baryonic mass distribution to be in much better agreement with observational constraints. Using X-ray data, it is indeed possible to estimate the gas profile in large clusters, while the stellar mass can be measured using optical data. \cite{Gonzalez:2007p916} have computed the baryon mass fraction within $r_{500}$ for a large sample of groups and clusters.  They have found a slight deficit of baryons $f_{\rm bar}=0.133 \pm 0.004$, when compared to the universal baryon fraction as measured by WMAP $\Omega_b/\Omega_m=0.176 \pm 0.008$ \citep{Spergel:2003p1726, Spergel:2007p1721}. From Figure~\ref{fig:BaryonFraction}, we see that the SF and Quenching runs show a slight baryon {\it excess} with $f_{\rm bar} \simeq 0.16$ while our universal baryon fraction $\Omega_b/\Omega_m$ was set to 0.15 in our simulation (see Table~\ref{tab:BaryonFraction}). This is a direct consequence of overcooling \citep{Kravtsov:2005p702}. On the contrary, the AGN model show a clear baryon {\it deficit} with $f_{\rm bar} \simeq 0.13$, in striking agreement with the observed average value. Note however that we used a universal baryon fraction lower than the WMAP estimate, so that we cannot claim that our model is fully consistent with the data yet.  Nevertheless, this baryon deficit in our model is the consequence of the repeated effect of AGN-driven shocks and convective motions, pushing gas outside the virial radius. One can see from the local baryon fraction profile in Figure~\ref{fig:BaryonFraction} that this gas accumulates in a region between 1 to 2 virial radii around the cluster, beyond which the cumulated baryon fraction converge to the universal one. 

Our SF model compares very well to the AMR simulation performed by \cite{Kravtsov:2005p702}, showing a slight excess of baryon at the virial radius. From these standard galaxy formation models, we obtain gas properties that compare favorably to X-ray data \citep{Nagai:2007p712}. We see indeed in Figure~\ref{fig:BaryonFraction} that in the SF model, the gas mass fraction is slowly decreasing towards the center, with a significant deficit at the virial radius. This behavior is traditionally explained by the joint effect of cooling and star formation \citep{Voit:2001p1750}. The price to pay is however to form too many stars and cold gas in the cluster, as confirmed by previous numerical models \citep{Borgani:2004p1066, Kravtsov:2005p702, Borgani:2009p728}. Our simple Quenching model is making things better for the stellar component, but now the gas mass distribution is too concentrated (see Fig.÷\ref{fig:BaryonFraction}). Only with AGN feedback can we obtain a small stellar mass fraction and in the same time a small gas fraction. We note however that, in our AGN model, the gas mass profile is not as steep as suggested by X-ray data. We could probably reproduce the gas profile of the SF model using another, more efficient AGN feedback scheme.

Recently, \cite{Puchwein:2010p763} have simulated  a large number of groups and clusters with the SPH code GADGET, using a mass resolution that is only about a factor 2 lower than ours and a spatial resolution of 2.5kpc (compared to 1 kpc at low res. or 0.5 kpc at high res. here). Nevertheless, they also found that with AGN feedback, the total baryon fraction was below the universal value. More interestingly, using a high universal baryon fraction ($\Omega_b/\Omega_m=0.165$), they report a stellar mass fraction of $f_* \simeq 0.05$, quite independent of the parent halo mass. In our case, for our Virgo-like cluster with $M_{\rm vir} \simeq 10^{14}$~M$_\odot$, we obtained $f_* \simeq 0.01$ in the low resolution case and $f_* \simeq 0.02$ in our high resolution simulation. Our value is in better agreement with the observational estimate proposed by \cite{Lin:2003p1820, Lin:2004p1817}, while the higher value found by  \cite{Puchwein:2010p763} is in better agreement with the observations reported in \cite{Gonzalez:2007p916}. Using the COSMOS survey, \cite{Giodini:2009p4283} have estimated stellar mass fractions for a large sample of galaxy clusters and groups. For our simulated halo mass, $M_{500c} \sim 8 \times 10^{13}$~M$_\odot$, they report stellar mass fraction ranging from 2\% to 5\%. Using the original feedback model of \cite{Booth:2009p501} in the GADGET code, \cite{Duffy:2010p568} have also computed the predicted baryon and stellar mass fraction of a large sample of groups extracted from a cosmological simulation. Although they report a similar baryon deficit within the virial radius, they obtained $f_* \simeq 0.03$ for an even larger universal baryon fraction $\Omega_b/\Omega_m=0.18$. We see that there is a consensus about a strong reduction of the stellar mass fraction in groups and clusters thanks to AGN feedback. The extent of this reduction seems to depend quite sensitively of the details of each implementation, and possibly to the nature of the code (SPH versus AMR). As suggested by observations also, we note that the exact stellar and baryon fraction probably varies from halo to halo. We would like also to stress that all the reported simulations, including ours, are probably not fully converged yet.

\section{Summary and Conclusions}
\label{sec:summary}

We have simulated the formation of a Virgo--sized galaxy cluster to study the effects of feedback on the overcooling problem.
The impact of AGN feedback on the distribution of the baryonic mass is strong,
and in good agreement with previous SPH simulations: star formation in massive galaxies is drastically reduced. At the same time, left-over gas is very efficiently removed from the core of the parent halos, where it would have otherwise accumulated. 
In order to quantify the effect of AGN feedback, we have run two other reference simulations: one model with only star formation and supernovae feedback (the standard scenario) and one model for which we have artificially prevented star formation to occur in massive enough spheroids (the quenching scenario). 

A detailed comparison of the three models clearly demonstrates that AGN feedback is needed to control star formation in the central BCG, but also to unbind the overcooling gas from the cluster core. We also clearly identify the effect of the baryon dynamics on the dark matter mass distribution on large scale. Interestingly enough, in case of AGN feedback, we observe  the adiabatic expansion of the dark halo, an effect well modeled by the AC theory of \cite{Gnedin:2004p569}. A comparison of our simulation results with observational data for Virgo and its central galaxy M87 rules out the standard model, but also the quenching model. On the contrary, our simulation with AGN feedback, although not fully converged yet, shows a much better agreement with M87 data in term of mass distribution. In particular, we obtain a significantly reduced baryon fraction within the virial radius, in agreement with observations compiled by \cite{Lin:2003p1820} and \cite{Gonzalez:2007p916}. We clearly identify in our simulation that gas is removed from the core of the cluster by convective motions and/or strong shocks, and accumulates in a region just outside the virial radius. When compared to gas profiles inferred from X-ray data, our AGN model produces too shallow gas distribution, suggesting that we probably need even more powerful feedback processes.

Our cluster formation simulations with AGN feedback have not fully converged yet - as we increase the resolution, we find a stellar mass profile for the BCG that is in better agreement with observations, but it is still too low by about a factor of two. We are still  missing the lowest mass galaxy population, which could provide the missing stellar mass in the central elliptical galaxy. We also note that, in the current picture, AGN feedback is a morphologically dependent process: it only directly affects galaxies with SMBHs, i.e. galaxies with a significant bulge/spheroid component. Higher resolution studies would be needed, in order to reliably model this distinction, so that star formation in disky galaxies is not artificially suppressed.

\section*{Acknowledgments}
We thank our anonymous referee for helpful suggestions that greatly improved the quality of the paper.
RT thanks Andrey Kravtsov for stimulating comments.
All simulations were performed on the Cray XT-5 cluster at CSCS, Manno, Switzerland.
We thank the CTS for supporting the astrophysical fluids program at the University of Z\"urich.


\bibliography{papers}


\appendix

\section{Adiabatic contraction model}

If one defines the initial radius of each dark matter shell as $r_i$ and its final, adiabatically contracted value $r_f$, \cite{Abadi:2009p531} have proposed to capture \cite{Gnedin:2004p569} model using the following simplified model 
\begin{equation}
\frac{r_f}{r_i}=1+\alpha \left( \frac{M_i}{M_f}-1\right) ~~~{\rm with}~~ \alpha \simeq 0.68{\rm .} 
\label{equ:Abadi}
\end{equation}
The original \cite{Blumenthal:1986p732} can be recovered using $\alpha =1$. The final cumulated mass distribution is computed using
\begin{equation}
M_f = M_{\rm dm}(r_f) + M_{\rm bar}(r_f) = f_{\rm dm} M_i(r_i)+M_{\rm bar}(r_f)
\end{equation}
where we have assumed that the initial dark matter mass is conserved during AC. 
The initial dark matter distribution is described using the analytical NFW profile
\begin{equation}
M_i(r_i)=M_{\rm 200}\frac{\log(1+x)-x/(1+x)}{\log(1+c)-c/(1+c)}
\end{equation}
where $x=r_i/r_s$ and $r_s=r_{\rm 200}/c$. $M_{\rm 200}$ is the total virial mass. For the baryonic distribution, we assume a constant surface density disc with size $r_d$ and mass $m_d$, so that
\begin{equation}
M_{\bar bar} (r_f)= m_d \left( \frac{r_f}{r_d} \right)^2
\label{equ:Baryons}
\end{equation}
The dark matter mass fraction is computed using $f_d=1-m_d/M_{\rm 200}$. The model we considered in Equation~\ref{equ:Baryons} for the baryonic mass distribution has been chosen that simple on purpose: inserting Equation~\ref{equ:Baryons} into the AC relation in Equation~\ref{equ:Abadi}, one clearly sees that we have to find the only real root of a third order polynomial equation with unknown $r_f/r_i$. This can be done quite easily with any root finder. In case the disc size is small enough (namely if $r_d \ll r_s$), the AC model is fully tractable analytically by noticing that for $x \ll 1$, one has $M_i \propto x^2$. We therefore have 
\begin{equation}
\frac{r_f}{r_i}=1+\alpha \left( \frac{M_i(r_i)}{f_d M_i(r_i) + m_d}-1\right) ~~~{\rm for}~~Ê r_f \ge r_d
\end{equation}
\begin{equation}
\frac{r_f}{r_i} \simeq {\rm constant} ~~~{\rm for}~~Ê r_f < r_d
\end{equation}
where the constant can be determined by continuity.

\label{lastpage}
\end{document}